\documentclass[a4paper]{article}

\pdfoutput=1 

\usepackage{silence}

\usepackage{jcappub} 
\usepackage{longtable}
\usepackage{epsfig}
\usepackage[section]{placeins}
\usepackage{float}
\usepackage{color}

\usepackage{lineno} 

\usepackage{units} 
\usepackage[load-configurations=astronomy]{siunitx}
\DeclareSIUnit\parsec{pc}
\DeclareSIUnit\hubble{\ensuremath{h}}
\usepackage[dvipsnames]{xcolor}
\newcommand\myshade{85}
\colorlet{mylinkcolor}{violet}
\colorlet{mycitecolor}{Aquamarine}
\colorlet{myurlcolor}{YellowOrange}

\hypersetup{
  linkcolor  = mylinkcolor,
  citecolor  = mycitecolor!\myshade!black,
  urlcolor   = myurlcolor!\myshade!black,
  colorlinks = true,
}
\PassOptionsToPackage{bookmarks=false}{hyperref}

\title{Constraints from Ly-$\alpha$ forests on non-thermal dark matter including resonantly-produced sterile neutrinos}

\author[a]{Julien Baur,}
\author[a]{Nathalie Palanque-Delabrouille,}
\author[a]{Christophe Y\`eche,}
\author[b]{Alexey Boyarsky,}
\author[c]{Oleg Ruchayskiy,}
\author[a]{\'Eric Armengaud,}
\author[d]{Julien Lesgourgues}

\emailAdd{julien.baur@cea.fr}

\affiliation[a]{CEA/Irfu, Universit\'e Paris-Saclay,  F-91191 Gif-sur-Yvette, France}
\affiliation[b]{Intituut-Lorentz, Leiden University, Niels Bohrweg 2, 2333 CA Leiden, The Netherlands}
\affiliation[c]{Discovery Center, Niels Bohr Institute, Blegdamsvej 17, 2100, Copenhagen, Denmark}
\affiliation[d]{Institute for Theoretical Particle Physics and Cosmology (TTK), RWTH Aachen University, D-52056 Aachen, Germany}

\date{Received xx; accepted xx}

\abstract{
We use the large BOSS DR9 sample of quasar spectra to constrain two cases of non-thermal dark matter models: cold-plus-warm dark matter (C+WDM) where the warm component is a thermal relic, and sterile neutrinos resonantly produced in the presence of a lepton asymmetry (RPSN). We establish constraints on the  thermal relic mass $m_{x}$ and its relative abundance $F_{\rm{wdm}}=\Omega_{\rm{wdm}}/\Omega_{\rm{dm}}$ using a suite of cosmological hydrodynamical simulations in 28 C+WDM configurations. We find that the $3\sigma$  bounds in the 
$m_{x} - F_{\rm{wdm}}$ parameter space approximately follow 
$F_{\rm{wdm}} \sim 0.35 (\mathrm{keV}/m_x)^{-1.37}$ from BOSS data alone. 
We also establish constraints on sterile neutrino mass and mixing angle by further producing the non-linear flux power spectrum of 8 RPSN models, where the input linear power spectrum is computed directly 
from the particles distribution functions. We find values of lepton asymmetries for which sterile neutrinos as light as $\sim 6.5~\mathrm{keV}$ (\textit{resp.} $3.5~\mathrm{keV}$) are consistent with BOSS data at the 
$2\sigma$ (\textit{resp.} $3\sigma$) level. These limits tighten by close to a factor of 2 for 
values of lepton asymmetries departing from those yielding the coolest distribution functions. 

Our Lyman-$\alpha$ forest bounds can be additionally strengthened if we include
higher-resolution data from XQ-100, HIRES and MIKE that allow us to probe smaller scales. At these scales, the measured flux power spectrum exhibits a suppression  
 that can be due to Doppler broadening, IGM pressure smoothing or free-streaming of WDM particles.
In order to distinguish between these mechanisms, thermal history at redshifts $z \ge 5$ should be determined. 
In the current work, we show that if one extrapolates temperatures from lower redshifts via broken power laws in $T_0$ and $\gamma$, then 
our $3\sigma$ C+WDM  bounds strengthen to $F_{\rm{wdm}} \sim 0.20  (\mathrm{keV}/m_x)^{-1.37}$, and  the lightest resonantly-produced sterile neutrinos consistent with our extended data set have masses of 
$\sim 7.0~\mathrm{keV}$ at the $3\sigma$ level. 
In particular, using dedicated hydrodynamical simulations, we show that a hypothetical 7 keV sterile neutrino produced in a lepton asymmetry of $\mathcal{L} = \vert n_{\nu_e} - n_{\bar{\nu}_e} \vert  / s ~=~ 8 \times 10^{-6}$ is consistent at $1.9~\sigma$ (\textit{resp.}  $3.1~\sigma$) with BOSS  (\textit{resp.} BOSS + higher-resolution) data,  for the thermal history models tested in this work. 
More information about the state of the IGM at redshifts 5--6 will allow one to conclude whether the small-scale suppression of the flux power spectrum is due to such sterile neutrino or to thermal effects. 

}

\begin{document}

\hypersetup{pageanchor=false}
\begin{titlepage}
\maketitle
\end{titlepage}

\hypersetup{pageanchor=true}

\section{Introduction}
\label{sec:intro}
It appears with increasingly robust evidence from gravitational lensing,
non-Newtonian galaxy rotation curves, baryon acoustic oscillations (BAO) and
the cosmic microwave background (CMB) that 80\% of all gravitating matter in
the Universe is dark. Even under the assumption that dark matter (DM) is an
additional component to baryonic matter, the standard model of particle
physics currently has no viable DM candidate, \textit{i.e.,} a massive,
neutral, long-lived particle insensitive to the electromagnetic
interaction. The discovery of neutrino flavor oscillations (see,
\textit{e.g.},~\cite{Strumia:2006db}) proved our current picture of the
standard model to be incomplete by providing direct evidence for non-zero
neutrino masses. Neutrinos are, however, too light to account for all of dark
matter. Therefore, a particle origin of dark matter necessarily
implies  extensions of the standard model of particle physics, many of which provide dark matter candidates
(for a review see, \textit{e.g.},~\cite{Feng:2010gw,Taoso:07}). One such candidate,
similar to neutrino dark matter while being free of its drawbacks, is the
\emph{right-handed} (\textit{a.k.a.} sterile) neutrino~\citep{DodelsonWidrow94}.
Right-handed neutrinos carry no lepton charge and only directly
couple (mix at quadratic level) to left-handed (active) neutrinos. As such,
sterile neutrinos never thermalize in the early Universe. Therefore their
number density is much lower than that of the cosmic neutrino
background. Nevertheless, enough may be produced to account for the observed
total matter density, and are thus an inherent DM
candidate~\citep{DodelsonWidrow94,ShiFuller99,Dolgov:2000ew,Abazajian2001,Asaka:2006nq,LaineMSM}.
Although produced while relativistic, these particles become non-relativistic deeply in
the radiation dominated epoch (RDE) --- as any dark matter should --- and therefore evade
the cosmic microwave background constraint on the effective number of stable,
relativistic species in thermal equilibrium in the early Universe,
$N_{\rm{eff}} = 2.99 \pm 0.20$ (from the combined TT+TE+EE+lowP Planck
analysis \cite{Planck2015}).

Such particles, becoming non-relativistic deep in the RDE, are generically known as \emph{warm dark matter} (WDM). WDM particles free-stream out of gravitational potential wells, delaying structure growth on scales below a characteristic distance $\lambda_{\rm{FSH}}$ that scales with the expansion rate $H=\dot{a}/a$.
WDM is expected to produce fewer dark matter halo satellites than CDM~\cite{Bode2000}. However, there is still no direct way to probe the dark matter distribution in the Local Group other than by gravitational lensing. The fewer-than-expected galactic satellite count around the Milky Way does not constitute conclusive evidence in favor of a warm dark matter since there are several well understood mechanisms which lead to poorly-populated dark matter haloes such as re-ionization and feedbacks from active galactic nuclei (AGNs) and supernov{\ae}~\citep{Bose_reio}. Incorporating these and other baryonic effects on sub-Mpc scales in N-body simulations can also account for the observed less cuspy inner density profiles of galaxy halos. Distinguishing cold from non-cold dark matter (NCDM) requires either directly probing the dark matter distribution in galaxy clusters such as with strong gravitational lensing~\citep{WvsCDM, SL_proj,Birrer:2017rpp}, or probing the scales at which WDM is expected to produce no structures beyond a certain mass. Lyman-$\alpha$ (Ly-$\alpha$) forests in the spectra of distant $z \in \left[ 2.1, 4.5 \right]$ quasars (QSO) probe the intergalactic $\sim 10^{0-2}~h^{-1}~\rm{Mpc}$ scales encompassed by the integrated free-streaming of keV particles while they remained relativistic. They are thus a formidable probe for establishing lower bounds on NCDM particle mass~\citep{Hansen:2001zv,VLH08a, VLH08b, SMT08, VBH08, VBH13,BLR09,BLR09letter,warmIGM,Baur16}. 

Since they mix with left-handed neutrinos, sterile neutrinos can be copiously produced in the early Universe ~\citep{DodelsonWidrow94,Dolgov:2000ew, Abazajian2001,Asaka:2006nq}. A net lepton asymmetry (an excess
of leptons over anti-leptons) present in the primordial plasma during the production epoch boosts active-sterile neutrino mixing and thus can account for the observed abundance of dark matter~\citep{ShiFuller99,Abazajian2001,LaineMSM}. This requires lepton asymmetries of several orders of magnitude larger than the baryon asymmetry of the Universe~\cite{LaineMSM} estimated from light element abundances from big bang nucleosynthesis.
These resonantly-produced sterile neutrinos (RPSN) are relevant in that they
have much cooler distribution functions than is assumed in the so-called
non-resonant production (NRP) mechanism, which occurs in the absence of lepton
asymmetry. Through their cooler distribution functions on the one hand and
their smaller mixing angles on the other hand, a RPSN of mass
$m_{\nu_s}^{\rm{rp}}$ is more compatible with clustering, Ly-$\alpha$ and
X-ray data than a NRP neutrino of the same mass
$m_{\nu_s}^{\rm{nrp}}$~\citep{BLR09letter}. Additional production mechanisms
exist (see~\cite{Adhikari:2016bei}) including production from the decay of a
scalar field~\citep{Shaposhnikov:2006xi,Kusenko:2006rh,Petraki:2007gq} and
diluted thermal overproduction~\cite{Asaka:2006ek,Bezrukov:2009th} to name
just a few. They all have in common a cooler momentum distribution than NRP neutrinos which translate into shorter free-streaming horizon scales. 

In this work we investigate several NCDM dark matter model, different from
thermal relic WDM. Only a handful of similar studies have been conducted thus
far. Ref.~\cite{BLR09} issued bounds on the cold+warm dark matter mixtures,
using SDSS-II data. Ref.~\citep{BLR09letter} have issued bounds on the mass of
RPSN by rescaling the Ly-$\alpha$ bounds on the mixture of cold+warm dark
matter~\cite{BLR09}. Several other works~\cite{warmIGM,Schneider16,Murgia2017} used the
bounds on thermal relic WDM and extrapolated the expected flux power spectrum in the Ly-$\alpha$ forest using a correction from the linear matter power spectrum (in~\cite{Murgia2017}, the matching was also performed at the level of the non-linear power spectrum obtained from DM-only simulations).

The current work issues bounds on the mass of RPSN using dedicated hydrodynamical simulations, a approach which had not been implemented directly before for Ly-$\alpha$ forest constraints. The paper is outlined as follows.
Sec.~\ref{sec:method} recaps how Ly-$\alpha$ forests are used to probe the $10^{0-2}~h^{-1}~\rm{Mpc}$ scales through the flux power spectrum with the help of state-of-the-art hydrodynamical simulations. We mention how NCDM cosmologies are incorporated in those simulations in Sec.~\ref{subsec:simu}. We present the Ly-$\alpha$ forest power spectrum constructed from 3 samples of QSOs, from the Sloan Digital Sky Survey (SDSS), the Very Large Telescope (VLT) and W.~M.~Keck Observatory, that enable us to constrain NCDM cosmologies in Sec.~\ref{subsec:data}.
Sec.~\ref{sec:ncdm} discusses  different neutrino production mechanisms and their impact on the initial matter power spectrum,  used in our simulations pipeline. We also investigate mixed cold plus warm dark matter (C+WDM) models in Sec.\ref{subsec:cwdm} and use them to put constraints on the mass and fraction of the warm component. We also use these C+WDM models as a means to cross-check our bounds on RPSN mass by issuing a mapping between the parameters involved in both cases. We recap our results in Sec.~\ref{sec:results} and conclude in Sec.~\ref{sec:conclusion}.


\section{Probing the power spectrum with Ly-$\alpha$ Forests of QSOs}
\label{sec:method}
Neutral Hydrogen in the intergalactic medium (IGM) constitutes a biased tracer for the (total) matter density fluctuations at $1 - 100~h^{-1}~\rm{Mpc}$ scales. It is present wherever photo-ionization by the ultraviolet (UV) background is balanced by electron recombination $\gamma + \rm{H} \rightleftharpoons e^{-} + \rm{H}^{+}$. The density of neutral Hydrogen (\textsc{Hi}) can be obtained by equating the interaction rates of both reactions and by setting global neutrality ($n_e = n_p$):
\begin{equation}
	n_{\rm{\textsc{Hi}}} = \left( \frac{\sigma_{\rm{comb}} \sqrt{2 k_b/m_e}}{c \sigma_{\rm{ion}}} \right) \frac{n_b^2 \sqrt{T}}{n_{\gamma_{\rm{UV}}}}
		\label{eq:Hydrogen}
\end{equation} where $\sigma_{\rm{comb, ion}}$ are respectively the recombination and photo-ionization cross-sections\footnote{It should be noted that the interaction rates involved are implicit functions of temperature $T$. The dependence on $T$ is rather $n_{\rm{\textsc{Hi}}} \propto T^{-0.7}$ than $n_{\rm{\textsc{Hi}}} \propto T^{0.5}$ as is implied in Eq.~\ref{eq:Hydrogen}.} and $n_{b, \gamma_{\rm{UV}}}$ the baryon and UV photon densities.

Photons of wavelength $\lambda = 121.6~\rm{nm}$ interacting with a neutral Hydrogen atom have a probability $\propto e^{- \tau}$ of exciting the bound electron from the fundamental to the first excited state, known as the Ly-$\alpha$ transition. The Ly-$\alpha$ optical depth in the IGM $\tau$ thus relates the observed and intrinsic flux from a background quasar with a Hydrogen-rich IGM in the foreground via
\begin{equation}
	\varphi_{\rm{obs}} \left(\lambda\right) = e^{- \tau (\lambda)} ~\varphi_{\rm{qso}} \left(\lambda\right)
		\label{eq:flux}
\end{equation} in the (rest-frame) wavelengths that span between the Ly-$\alpha$ $\lambda$1216 and Ly-$\beta$ $\lambda$1026 emission lines. The optical depth is given by the Hydrogen density times the Ly-$\alpha$ absorption cross section $\sigma_{\rm{Ly}\alpha}$ integrated over distance. Since we probe density fluctuations in the Hubble flow, we measure quantities in velocity-space and thus the Ly-$\alpha$ optical depth at velocity $v$ (with respect to Earth at $v=0$) is the aforementioned quantity integrated along the line-of-sight 
\begin{equation}
	\tau \left(v\right) = \int_{0}^{v} \mathrm{d}v'_{\parallel} ~ \frac{n_{\rm{H_I}} \sigma_{\rm{Ly}\alpha}\left(z\right)}{\nabla v'_{\parallel}}
		\label{eq:tau}
\end{equation} where $\nabla v_{\parallel}$ is the velocity gradient parallel to the line-of-sight. \\

Because of cosmological redshift, the rest-frame wavelength of the Ly-$\alpha$ absorption gets redshifted from its observed wavelength as the quasar signal travels to Earth in an expanding Universe. Hence a series of absorption features  between the Ly-$\alpha$ and Ly-$\beta$ emission lines in the spectra of high-redshift QSOs that entails the distribution of neutral Hydrogen along the line-of-sight. This Ly-$\alpha$ forest is a widely-used tool to probe density fluctuations at intergalactic scales. \\

\subsection{From Ly-$\alpha$ forest to the flux power spectrum}
\label{subsec:data}

\begin{figure}
\begin{center}
	\includegraphics[width=\columnwidth]{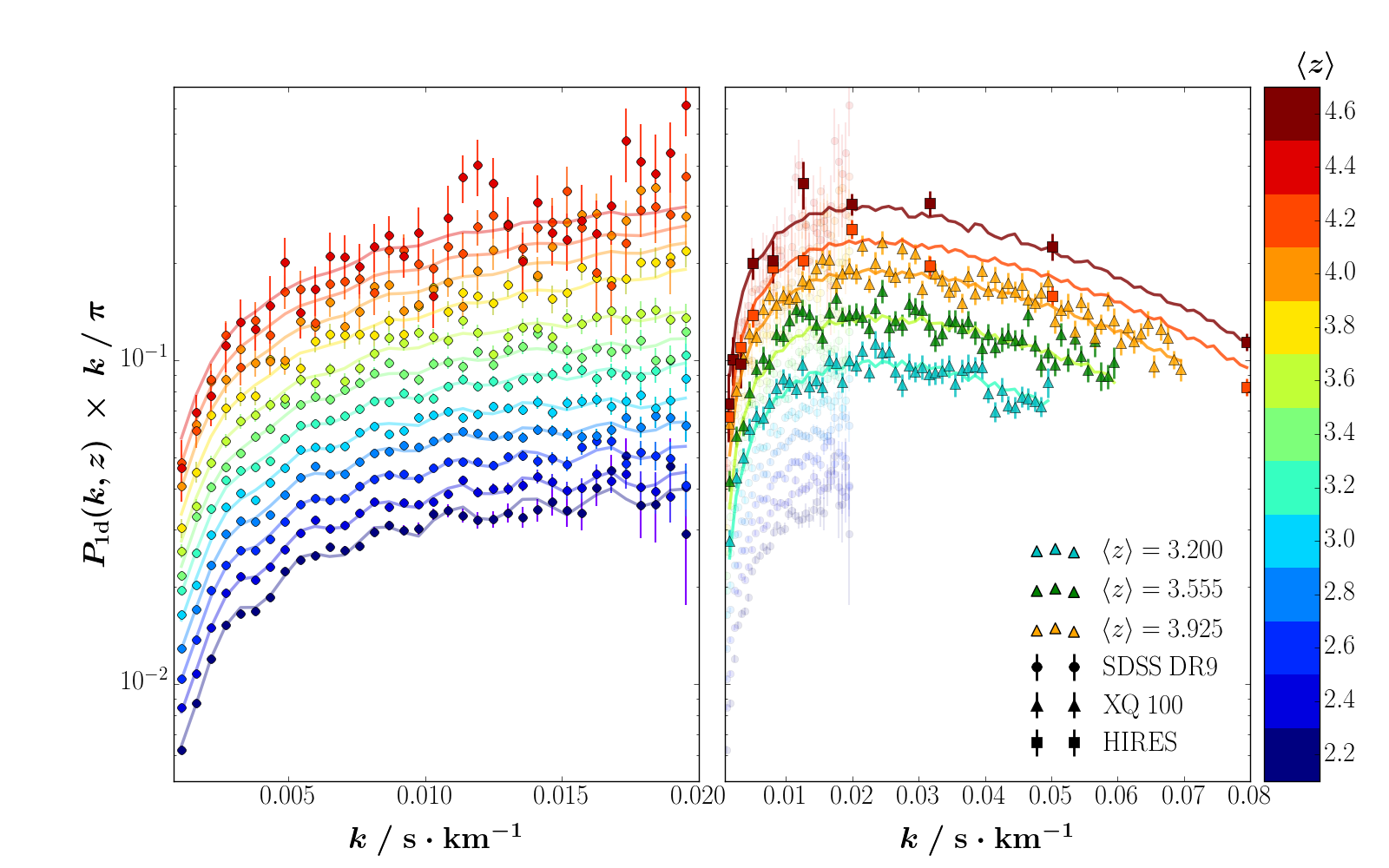}
	\caption{Dimensionless Ly-$\alpha$ flux power spectra $\Delta^2_{\varphi} (k) = P_{\varphi} (k) \times k/\pi$ from our selected samples. Color encodes redshift bin. Solid lines are our best fit in each redshift bin (see Sec.~\ref{subsec:res_cwdm}). The apparent oscillations arise from the Ly-$\alpha$--SiIII correlation which occurs at a $\Delta \lambda = 9.2~\rm{\AA}$ separation. \textbf{Left:} BOSS DR9 data only. \textbf{Right:} Three redshift bins of the XQ-100 sample are displayed with the nearest corresponding simulation ($z=3.2, 3.6, 4.0$), along with the two lowest redshift bins of the HIRES/MIKE sample ($z=4.2,4.6$).
}
		\label{fig:DR9fPS}
\end{center}
\end{figure}

Perhaps the most convenient and widely-used statistical tool to compare observations with theoretical predictions is the one-dimensional flux power spectrum~\citep{Croft1999}. It is obtained by the Fourier transform of the QSO's transmitted flux fraction
\begin{equation}
	\delta_\varphi (\lambda) = \frac{\varphi (\lambda) - \langle \varphi \rangle}{\langle \varphi \rangle} = \frac{e^{-\tau}}{e^{-\tau_{\rm{eff}}}} - 1
		\label{eq:delta}
\end{equation} normalized by a mean transmitted flux which defines an effective optical depth $\langle \varphi \rangle = e^{- \tau_{\rm{eff}}}$. \\

The flux power spectrum in the Ly-$\alpha$ forest is obtained by deconvolving the power spectrum of the transmitted flux fraction defined in Eq.~\ref{eq:delta} from the instrumental window function $W$ after substracting the (white) noise power spectrum, and averaging over the entire set of quasars:
\begin{equation}
	P_{\rm{Ly}\alpha} (k) = \left\langle \frac{\vert \tilde{\delta_\varphi}(k) \vert^2 - P_{\rm{noise}}(k)}{W^2 (k)} \right\rangle_{QSOs}
		\label{eq:PLya}
\end{equation} 

\subsubsection{SDSS / BOSS}

Our sample consists of $13,821$ out of a parent sample of $\sim 60,000$ quasar spectra from the DR9 of SDSS-III/BOSS~\citep{Ahn2012, Dawson2012, Eisenstein2011, Gunn2006, Ross2012,Smee2013}. They are selected for the following criteria on their Ly-$\alpha$ forest: a signal-to-noise ratio per $\Delta \lambda / \lambda = 10^{-4}$ pixel greater than 2,  absence of broad absorption line features, absence of damped or detectable Lyman-limit systems, and  an average resolution in the Ly-$\alpha$ forest of at most $85~\rm{km}~s^{-1}$.\\

The spectra in this sample are used to  measure the transmitted flux power spectrum in 12 redshift bins from $\langle z \rangle = 4.4$ to $2.2$ (each bin spanning $\Delta z = 0.2$) and in 35 equally-spaced spatial modes  ranging from $k=10^{-3}$ to $2.10^{-2}~s~\rm{km}^{-1}$ (\textit{cf.} left panel of Fig.~\ref{fig:DR9fPS}). To reduce correlations between neighboring $z$-bins, we split the Ly-$\alpha$ forest of each quasar spectrum into up to three distinct redshift sectors. Each sector has a maximum extent of $\Delta z < 0.2$. The transmitted flux power spectrum $\vert \tilde{\delta_\varphi}(k) \vert^2$ is computed separately in each $z$-sector. We checked that the resulting power spectrum agreed with that derived from a likelihood approach.\\

Complete details on our selection procedure as well as calibrations, computation of the flux power spectrum and  determination of both statistical and systematic uncertainties are extensively described in~\cite{Palanque-Delabrouille2013}.

\subsubsection{VLT / X Shooter}

In addition to the SDSS/BOSS data described above, we make use of the Ly-$\alpha$ forest power spectrum from the XQ-100 Legacy Survey \citep{XQ100}, which consists of a sample of 100 medium-resolution QSO spectra observed with the VLT/XShooter instrument~\cite{XShooter}. We measure the Ly-$\alpha$ power spectra in 70 $k$ bins for 3 redshifts centered on $\langle z \rangle = 3.200, 3.555, 3.925$ following the methodology described in \cite{Yeche17} and shown in the right panel of Fig.~\ref{fig:DR9fPS}. We overlay the power spectra constructed with our hydrodynamical simulations in their central configuration (see Sec.~\ref{subsec:simu}) extracted from the 3 nearest redshift bins ($z=3.2, 3.6, 4.0$). The raw power spectrum measured in these bins asymptotically approaches the noise power spectrum computed as a white noise at small scales. Because of an uncertainty on the correction of the spectrograph resolution, we limit our study to 50, 60 and 70 $k$ bins respectively (corresponding to $k \leq 0.05, 0.06, 0.07~s~\mathrm{km}^{-1}$) for the 3 aforementionned redshifts. This screening ensures the raw power spectrum is always dominant over the noise power spectrum at these small scales.

\subsubsection{HIRES and MIKE}

Finally, we add the Ly-$\alpha$ power spectrum measured by \cite{HRdata} in high-resolution QSO spectra taken with the Magellan Inamori Kyocera Echelle (MIKE \cite{MIKE}) instrument at the Las Campanas Observatory and the High Resolution Echelle Spectrometer (HIRES \cite{HIRES}) at the W.M. Keck Observatory. This high-resolution power spectrum is measured in 4 redshift bins ($\langle z \rangle = 4.2, 4.6, 5.0, 5.4$) and 9 $k$ bins spanning down to $k \leq 0.08~ s~\mathrm{km}^{-1}$. Since we extract our simulated power spectra from $z=4.6$ downwards, we only make use of the two lower $z$ bins for this data set, which we plot on the right-hand panel of Fig.~\ref{fig:DR9fPS} along with the simulated power spectrum extracted at the corresponding redshifts. \\

Our combined Ly-$\alpha$ power spectrum data thus consist of $35 \times 12$ high-statistic low-resolution measurements with BOSS, $50+60+70$ low-statistic medium-resolution measurements with XQ-100, and $9 \times 2$ low-statistic high-resolution measurements with HIRES/MIKE for a total of 618 measurements that span various redshifts and scales.  

\subsection{Probing the non-linear power spectrum at galactic scales}
\label{subsec:simu}

We compare the Ly-$\alpha$ flux power spectrum ($P_\varphi (k)$ herein) described in~\cite{Palanque-Delabrouille2013} and in Sec.~\ref{subsec:data} above with a prediction model obtained by simulating the distribution of matter in the photo-ionized IGM using the hydrodynamics code \textsf{Gadget-3}\footnote{\tt http://www.mpa-garching.mpg.de/gadget/}~\citep{Springel2001, Springel2005}. We extract the position, density and velocity fields of a baryon and a DM population in 13 equidistant $\Delta z = 0.2$ redshift snapshots in $z \in \left[ 2.2, 4.6 \right]$. In each of these redshift bins, we compute the Ly-$\alpha$ optical depth from the underlying Hydrogen density distribution and IGM temperature (see Eq.~\ref{eq:Hydrogen}) from a sample of $10^{5}$ randomly traced lines of sight. The IGM temperature is extracted from a sample of $10^{6}$ particles from which we derive the temperature-density power law slope and intercept:
\begin{equation}
	T(\delta, z) = T_0 (z) \times (1+\delta)^{\gamma(z)-1}
		\label{eq:IGM}
\end{equation} The power spectrum of neutral Hydrogen density fluctuations in this velocity field is derived from these computed quantities. The photo-ionization rate in the IGM is fixed in all 13 redshift bins to be in agreement with typical measurements (see for instance \cite{Meiksin2009}), which is equivalent to re-normalizing the $P_\varphi (k)$ amplitude through the effective optical depth, which we model by a simple redshift power law:
\begin{equation}
	- \ln \langle \varphi \rangle = \tau_{\rm{eff}} = A^\tau \times (1+z)^{\eta^\tau}
		\label{eq:UV}
\end{equation} 
We derive the $P_\varphi (k)$ of $N^3=3072^3$ particles (for each species) in a $L^3 = (100~\rm{Mpc})^3$ co-volume with periodic boundary conditions from a subset of 3 smaller simulations. We correct for their lack of resolution or size using a splicing technique described in~\cite{McDonald2003}. Residuals between the exact and the spliced $P_\varphi (k)$ have been tested and modelled using $N=$ 1024, 1600 and 2048 simulations~\citep{Palanque2015b}. The simulations were run at the French TGCC\footnote{Tr\`es Grand Centre de Calcul} supercluster \textsf{Curie} under three PRACE\footnote{Partnership for Advanced Computing in Europe} and a GENCI\footnote{Grand \'Equipement National de Calcul Intensif} allocations totalling $19 \times 10^6$ CPU hours (see Acknowledgements). \\

To derive constraints on cosmological parameters, we compute the $P_\varphi (k)$ from a central benchmark (``best-guess" herein) model which sets our chosen free parameters at their central values in Tab.~\ref{tab:params}. Our cosmological parameters ($h$, $\Omega_m$, $\sigma_8$, $n_s$) are centered on the Planck 2013~\citep{PlanckCollaboration2013} best fitted values. The baryon density parameter $\omega_b = \Omega_b h^2$ has a negligible impact on the flux power spectrum, and is therefore kept fixed to the Planck 2013 best-fit value $\omega_b=0.04858$. Astrophysical parameters ($T_0^{z=3}$, $\gamma^{z=3}$, $A^\tau$, $\eta^\tau$) describing IGM thermodynamics\footnote{the $T_0$ and $\gamma$ mentionned here are the logarithmic $y$-intercepts of a redshift-dependant power law for $T(z),\gamma(z) ~ \propto \left[ (1+z)/4 \right]^\eta$, taken at a pivot $z=3$. The logarithmic slopes $\eta_{\gamma}$, $\eta_{T}(z<3)$ and $\eta_{T}(z>3)$ are let free and fitted in addition to $T_0 (z=3)$ and $\gamma (z=3)$.} (Eqs.~\ref{eq:IGM},\ref{eq:UV}) are consistent with measurements in~\cite{Schaye2000, Lidz2010, Becker2011, Garzilli2012, Meiksin2009}. We then compute the first and second order derivatives of $P_\varphi (k)$ and use them to derive a second-order Taylor expansion around our \textit{best-guess} model, with our derivatives taken at the steps in Tab.~\ref{tab:params}. \\
\begin{table}
	\begin{center}
		\begin{tabular}{ll}
			\textbf{parameter} & \textbf{central value \& step} \\[2pt]
			\hline \\[-10pt]
			$H_0$ & $(0.675 \pm 0.05) \times 10^2~\mathrm{km}~s^{-1}~\mathrm{Mpc}^{-1}$ \\[2pt]
			$\Omega_m$ &  $0.31 \pm 0.05$ \\[2pt]
			$n_s$ &  $0.96 \pm 0.05$ \\[2pt]
			$\sigma_8$ &  $0.83 \pm 0.05$ \\[2pt]
			$T_0^{z=3}$ &  $(1.4 \pm 0.7) \times 10^{4}~\rm{K}$ \\[2pt]
			$\gamma^{z=3}$ &  $1.3 \pm 0.3$ \\[2pt]
			$A^\tau$ &  $(2.5 \pm 2.0) \times 10^{-3}$ \\[2pt]
			$\eta^\tau$ &  $3.7 \pm 0.4$ \\[2pt]
			\hline \\[-10pt]
		\end{tabular}
	\end{center}
	\caption{Hydrodynamical simulation parameter grid. Our \textit{best-guess} is run on the central values. First and second order derivatives are computed using the stepped values in the right column.}
	\label{tab:params}
\end{table}

The \textsf{Gadget} code solves for the hydrodynamics at the onset of the non-linear regime of structure formation, which we arbitrarily set at $z=30$. We solve the Boltzmann equations in the expanding Universe numerically with the \textsf{CAMB}\footnote{\tt http://camb.info} \citep{Lewis2000}  and \textsf{CLASS}\footnote{\tt http://class-code.net/} \citep{CLASS} softwares and compute the linear power spectrum of matter density fluctuations at $z=30$. When implementing different NCDM cosmologies, it is useful to define a linear transfer function $T(k)$ defined such that $P_{\rm{ncdm}}(k) = T^2 (k) P_{\rm{cdm}}(k)$, where
\begin{equation}
	P_{\rm{cdm}}(k,z) = \mathcal{T}^2 (k,z) \times \mathcal{P} (k) \times \mathcal{D}^2(z)
		\label{eq:P3Dcdm}
\end{equation} is the linear (three-dimensional) power spectrum of total (baryon+DM) matter density fluctuations in the benchmark $\Lambda$CDM model and 
\begin{equation}
	\mathcal{P} (k) = \frac{2 \pi^2}{k^3} \mathcal{A}_s \left( \frac{k}{k_\star} \right)^{n_s - 1}
		\label{eq:primordial}
\end{equation} is the primordial scalar power spectrum. Because the Ly-$\alpha$ flux power spectrum is a unidimensional probe, we make the distinction between 1D and 3D in the above definition of the transfer function, where 
\begin{equation}
	2 \pi P^{\rm{1d}}(k) = \int_{k}^\infty dk' k' P^{\rm{3d}}(k')
		\label{eq:1D}
\end{equation}

A complete description of our simulations and its technical aspects are extensively detailed in~\cite{Borde2014}. The implementation of our different NCDM cosmologies is detailed in Sec.~\ref{sec:ncdm}.

\subsection{Systematics and caveats}
\label{subsec:caveats}

Although the Ly-$\alpha$ forest is a formidable albeit biased tracer for the matter distribution at intergalactic scales, several factors limit the accuracy of measurement and modeling of the $P_\varphi (k)$. These include feedback processes, such as active galactic nuclei (AGN) and supernov{\ae}, which can potentially correlate very disparate scales. Other baryonic effects can intervene at small scales~\citep{Semboloni:2011fe,vanDaalen:2011xb,Feedbacks, warmIGM, UVbackground}. Moreover, the scarce existing measurements of the IGM thermal state and  UV photo-ionization background at the relevant $z \geq 3$ redshifts compel us to have a very conservative apprehension of the astrophysical parameters. We therefore model the redshift dependence of the IGM density-temperature normalization and power index  as simple power-laws and we use large step sizes in the computation of the first and second order derivatives for the Taylor expansion, in order to encompass the large observational uncertainties. 
We also introduce in the fitting procedure several nuisance parameters which account for uncertainties related to our estimate of the instrumental noise in the data or of the spectrograph resolution, to our modeling of the IGM, to residual biases in the splicing technique used to achieve high-resolution simulations over 100~$h^{-1}\,\rm Mpc$ scales, to supernova and AGN feedbacks, and to the redshift of reionization. Details on the scale and redshift dependence of the nuisance parameters are extensively described in~\cite{Palanque2015a, Palanque2015b}.

It should be noted that at the velocity scales probed by the medium-resolution
BOSS DR9 data ($k \leq 2 \times 10^{-2}~s~\rm{km}^{-1}$), these small-scale
baryonic effects and the warmth of the IGM are not a limiting factor. For this
reason, throughout this paper we quote bounds on NCDM masses using BOSS data only ('SDSS'), in addition to our full set of BOSS + XShooter + MIKE + HIRES data ('SDSS+XQ+HR').


\section{NCDM signatures on the matter power spectrum}
\label{sec:ncdm}

When particles have relativistic enough velocities, they  free-stream to a horizon scale $\lambda_{\rm{FSH}}$ effectively unaffected by gravitational potentials. Hence the matter power spectrum is suppressed below the free-streaming horizon, which is given by 
\begin{equation}
\lambda_{\rm{FSH}}(t) = a(t) \int_0^{a(t)} da \frac{\langle v \rangle}{a^2 H}
\end{equation}
where the velocity dispersion $\langle v \rangle$ is given by the speed of light during the relativistic regime, and by  ${\langle p \rangle}/{m}$ afterwards. For warm and cool dark matter, this transition takes place during the radiation dominated era. The associated comoving scale
\begin{equation}
\frac{\lambda_{\rm{FSH}}(t)}{a(t)} = \frac{2 \pi}{k_{\rm{FSH}}(t)}
\end{equation}
grows like $t^{1/2}$ during the relativistic regime, like $\ln(t)$ during the non-relativistic regime as long as radiation dominates, and remains asymptotically constant during matter domination. Finally the comoving free-streaming horizon today can be estimated from
\begin{equation}
\frac{\lambda_{\rm{FSH}}^0}{a_0} = \frac{2 \pi}{k_{\rm{FSH}}^0}
\simeq \int_0^{a_{\rm{nr}}} \frac{da}{a^2 H} +  \int_{a_{\rm{nr}}}^{a_0} \frac{a_{\rm{nr}} da}{a^3 H}~,
\label{eq:FSH_L}
\end{equation}
where $a_{\rm{nr}} \simeq {\langle p \rangle_0}/{m}$ is the scale factor at
the time of the non-relativistic transition, and $\langle p \rangle_0$ is the
momentum dispersion today. Distant QSOs probe the power spectrum at scales of several Mpc (see Sec~\ref{subsec:data}). They enable putting
upper bounds on $k_{\rm{FSH}}^0$ of keV NCDM particles, which translate into
lower bounds on their mass. The velocity and momentum dispersion requires
knowledge of the explicit distribution function of the particle, which differs
from a Fermi-Dirac (thermal) distribution function depending on the production mechanism. Thus \emph{the mass bounds are different for each production mechanism}. 

Below we discuss several non-CDM (NCDM) models, inspired by the
sterile neutrino dark matter, that are the subject
of the current investigation.

\subsection{Non-resonantly produced sterile neutrinos as WDM}
\label{subsec:wdm}

Sterile neutrino were originally proposed as dark matter candidates by Dodelson and
Widrow~\cite{DodelsonWidrow94} (DW herein).  In the framework of DW,
sterile neutrinos are predominantly produced  at 
$T \sim 150~\mathrm{MeV}(m_{\nu_s}/\unit{keV})^{1/3}$ when the oscillation production
rate is most efficient while not reaching thermal
equilibrium~\cite{DodelsonWidrow94,Dolgov:2000ew,Abazajian:2001nj,Asaka:2006nq}.  The resulting distribution
function can be roughly approximated by a rescaled Fermi-Dirac
distribution~\cite{Dolgov:2000ew}, in which case the average momentum
$\langle p\rangle$ would be identical to that of active neutrinos. The
proper treatment however, based on the quantum Liouville equation~\cite{Asaka:2006rw}
shows that sterile neutrinos produced in (non-resonant) oscillations do not
feature a re-scaled thermal distribution and their average momentum is about 10--40\% colder
depending on sterile neutrino mass (see Fig.~8 in~\cite{Asaka:2006nq} or
Fig.~6 in~\cite{Laine:2008pg}). To distinguish them from the idealized DW
case, we refer to sterile neutrinos produced via this mechanism as \emph{non-resonant} (see Sec.~\ref{subsec:rpsn} for distinction), or NRP sterile neutrinos.

The requirement  $\Omega_{\nu_s} = \Omega_{\rm{dm}}$ fixes the $\theta$ --
$m_{\nu_s}^{\rm{nrp}}$ relationship, represented as the upper black solid line
in Fig.~\ref{fig:RPSN_MT}. The flux of photons from the radiative decay channel
$\nu_s \rightarrow \gamma \nu_\alpha$ is a function of $\theta$ and
$m_{\nu_s}$~\cite{Pal:1981rm}. Decay lines in astrophysical spectra, or the
lack thereof, thus establishes constraints on these
parameters~\cite{Dolgov:2000ew,Abazajian:2001vt,Boyarsky:2005us,Boyarsky:2006fg}. Comparing
the upper bounds on the putative dark matter decay flux with currently measured  DM abundance, \cite{BNRST06} and \cite{BNR07} yield an upper limit of $m_{\nu_s}^{\rm{nrp}} \leq 4~\rm{keV}$ for the NRP mechanism. The non-detection of small-scale damping in the flux power spectrum of the Ly-$\alpha$ forest due to $\nu_s$ free-streaming has yielded lower bounds consistently above the $4~\rm{keV}$ limit with $\geq 5 \sigma$ (see Tab.~\ref{tab:Lyabounds} below, with first constraints dating back to 2006~\cite{SMT08}). If right-handed neutrinos constitute all of dark matter, a growing consensus suggests they cannot be produced in this oscillation mechanism in absence of a net lepton asymmetry.

\begin{table*}
	\begin{center}
	\begin{small}
		\begin{tabular}{llllll}
			\textbf{Reference} & \textbf{QSO spectra} & \textbf{Data} & \textbf{Simultion} & \textbf{Bounds on $m_{\nu_s}^{\rm{nrp}}$} & \textbf{Tension with}\\[2pt]
			\textbf{study} & \textbf{resolution} & \textbf{set} & \textbf{resolution} & \textbf{from Ly-$\alpha$ forests} & \textbf{X-ray bounds}\\[2pt]
			\hline \\[-10pt]
			\cite{SMT08} & low only & SDSS-I & 12.8 (hydro)& $\geq~12$ keV & $6~\sigma$\\[2pt]
			\cite{BLR09} & low + high & SDSS-I + UVES & 6.7 (N-body) & $\geq~10$ keV& $5~\sigma$\\[2pt]
			\cite{VBH13} & high only & HIRES + MIKE & 25.6 (hydro) & $\geq~19$ keV& $9.5~\sigma$\\[2pt]
			\cite{Baur16} & low only & SDSS-III & 30.72 (hydro) & $\geq~24$ keV& $12~\sigma$\\[2pt]
			\cite{Yeche17} & low + medium & SDSS-III + XQ100 & 30.72 (hydro) & $\geq~25$ keV& $12.5~\sigma$\\[2pt]
			\cite{IrsicWDM} & medium + high & XQ100 + MIKE & 38.4 (hydro) & $\geq~34$ keV& $17~\sigma$\\[2pt]
			\hline \\[-10pt]
		\end{tabular}
	\end{small}
	\end{center}
	\caption{Summary of Ly-$\alpha$ constraints on NRP neutrino mass
          according to the data set used. Quoted lower bounds are the 95\%
          confidence level. Tension with the upper bound from X-rays is
          expressed in standard deviations in the right-most column.
          Simulation resolution refers to the quantity $\frac{N}{L }$ in $ \rm{Mpc^{-1}}$.}
	\label{tab:Lyabounds}
\end{table*}

\subsection{Mixed C+WDM models}
\label{subsec:cwdm}

\begin{figure}
\begin{center}
\includegraphics[width=0.6\columnwidth]{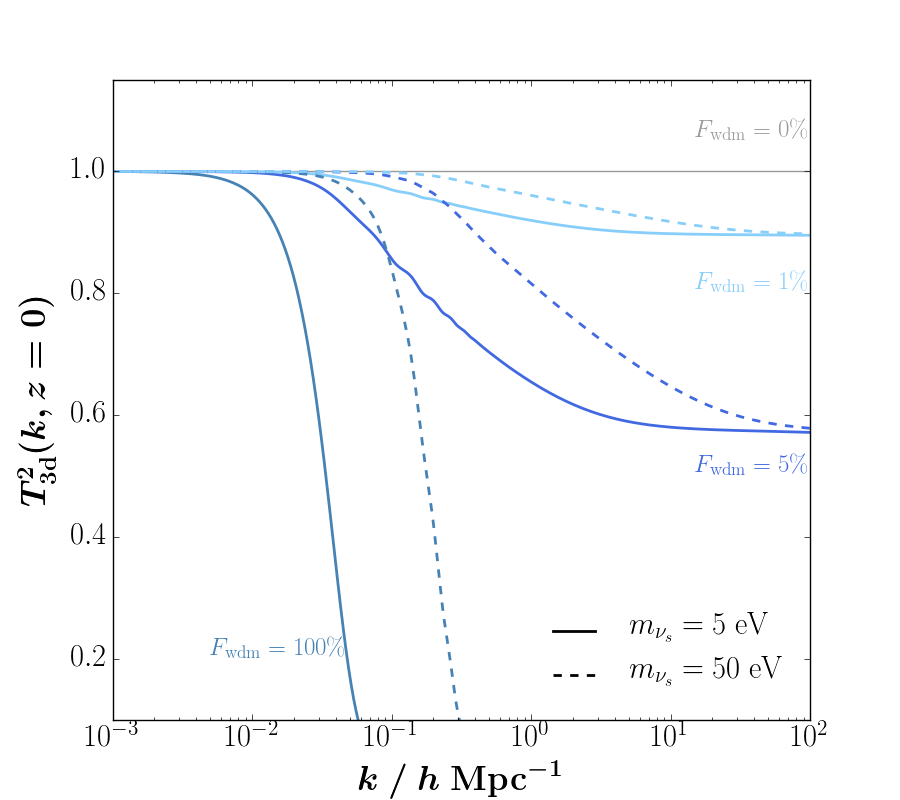}
\caption{3D linear transfer function for total matter at $z=0$ computed by the \textsf{CLASS} software for $m_{\nu_s} = 5~\rm{eV}$ in solid lines and $m_{\nu_s} = 50~\rm{eV}$ in dashed lines. Dark teal lines assume a pure WDM model, and lighter tones of blue display a larger preponderance of the cold (heavier) component over the warm (lighter) one.}
\label{fig:T3D_cwdm}
\end{center}
\end{figure}

A wide class of NCDM dark matter models can be approximated by adding a CDM component in addition to, \textit{e.g.}, thermal WDM. Such models start to deviate from CDM at scales determined by the mass of WDM component, but the overall amount of suppression is controlled by the 
warm DM fraction, $F_{\rm{wdm}}$, of the total dark matter. The warm-to-total DM fraction can thus be defined such that
\begin{equation}
\Omega_{\rm{wdm}} = F_{\rm{wdm}} \times \Omega_{\rm{dm}} = F_{\rm{wdm}} \times \left( \Omega_{\rm{wdm}} + \Omega_{\rm{cdm}} \right) 
\label{eq:Fwdm}
\end{equation}
The solid and dashed transfer functions ($T(k) = \sqrt{P_{\rm{ncdm}}/P_{\rm{cdm}}}(k)$) at $z=0$ in Fig.~\ref{fig:T3D_cwdm} feature the free-streaming cutoff scale for different NCDM masses. As discussed above, heavier DM particles damp power on smaller scales, which makes them more consistent with the benchmark $\Lambda$CDM model, at least in the linear regime.
In a cold plus warm dark matter model (C+WDM), the smaller is the fraction of the warm component, the colder is the overall transfer function as is shown by the lighter colored lines in Fig.~\ref{fig:T3D_cwdm} reaching an asymptotical plateau when $k \rightarrow \infty$ whose height is a function of the warm-to-cold fraction. For low $F_{\mathrm{wdm}} \lesssim 5\%$ ratios, the plateau is well approximated by $1-T(k \rightarrow \infty) \sim (1-F_{\mathrm{wdm}})$ (see \cite{BLR09}). For masses of a few keV, the thermal velocities of WDM particles can be neglected and thus the total dark matter distribution can be treated as a mono-species collisionless fluid in our hydrodynamical simulations whose linear transfer function is obtained by setting the DM abundance as $F_{\rm{wdm}}$ warm and $1-F_{\rm{wdm}}$ cold. The warm-to-total DM fraction $0 \leq F_{\rm{wdm}} \leq 1$ is therefore an additional free parameter that interpolates between the pure WDM limit ($F_{\rm{wdm}} = 1$) described above and the benchmark CDM ($F_{\rm{wdm}} = 0$) limit.

We run 28 C+WDM models on a non-uniform ($\mathrm{keV}/m_x$, $F_{\rm{wdm}}$) grid referenced as the black dots in Fig.~\ref{fig:Chi2_cwdm}, where $m_x$ is the mass of the WDM component assuming it is an early-decoupled thermal relic. NRP sterile neutrinos and thermal relics feature some nearly identical transfer functions, such that one can establish a correspondance between $m_x$ and $m_{\nu_s}$ from matching their relic density and average velocity. This gives the following mass mapping:
\begin{equation}
m_{\nu_s} = \kappa ~ m_x^{\mu} / \omega_{\rm{wdm}}^{1/3}
\label{eq:mxms}
\end{equation} where $\kappa = 4.43~\rm{keV}$, $\mu = 4/3$, and $\omega_{\rm{wdm}} = F_{\rm{wdm}} \times \Omega_{\rm{dm}}h^2$ is expressed in units of $0.25 \times 0.7^2$~\cite{VLH08a}. In reality,  this mapping is only approximate, since the actual phase-space distribution of NRP sterile neutrinos departs slightly from a rescaled Fermi-Dirac distribution~\cite{Asaka:2006rw,Asaka:2006nq}, and the authors of \cite{Abazajian2016} suggest a mapping with $\kappa = 3.90~\rm{keV}$ and $\mu = 1.294$. 
Since our set of hydrodynamical simulations used to establish constraints on NRP neutrinos as pure WDM were run with $\mathrm{keV}/m_x$ as the variable, we chose to do the same for our set of C+WDM models for the sake of consistency. Sterile neutrino masses can be established simply by using the mapping in Eq.~\eqref{eq:mxms} without running additional simulations. 

\subsection{Resonantly-produced Sterile Neutrinos (RPSN) as Cool DM}
\label{subsec:rpsn}

The presence of a net lepton asymmetry at temperatures $T \sim 0.1~\rm{GeV}$
can significantly enhance the production of sterile neutrinos from active
neutrinos through forward scattering in dense media \cite{ShiFuller99}. In a mechanism similar to the MSW\footnote{Mikheyev-Smirnov-Wolfenstein} effect~\cite{Mikheev:1986gs,Wolfenstein:1977ue} accounting for the solar neutrino deficit, the excessive abundance of leptons and their conjugate neutrinos with respect to anti-leptons can yield the correct DM density $\Omega_{\rm{dm}}$ with weaker mixing angles $\theta$. The authors of \cite{ShiFuller99} showed that this resonant production (RP) yields sterile neutrinos with significantly cooler momenta than the NRP ones. The resonant momenta depend on the sterile neutrino mass $m_{\nu_s}^{\rm{rp}}$ and net leptonic (assumed electronic) asymmetry $\mathcal{L} = (n_{\nu_e} - n_{\bar{\nu}_e}) / s$ in units of entropy density where $s \propto g_\star T^3$. If the resonance occurs before the QCD phase transition, only the low momenta states are populated from the quasi thermally-distributed active neutrinos ($\langle q = p/T_\nu \rangle \simeq 3.15$), resulting in cooler neutrino and anti-neutrino distribution functions with $\langle q \rangle \simeq 1.6$.

\begin{figure*}
\begin{center}
\includegraphics[width=16cm]{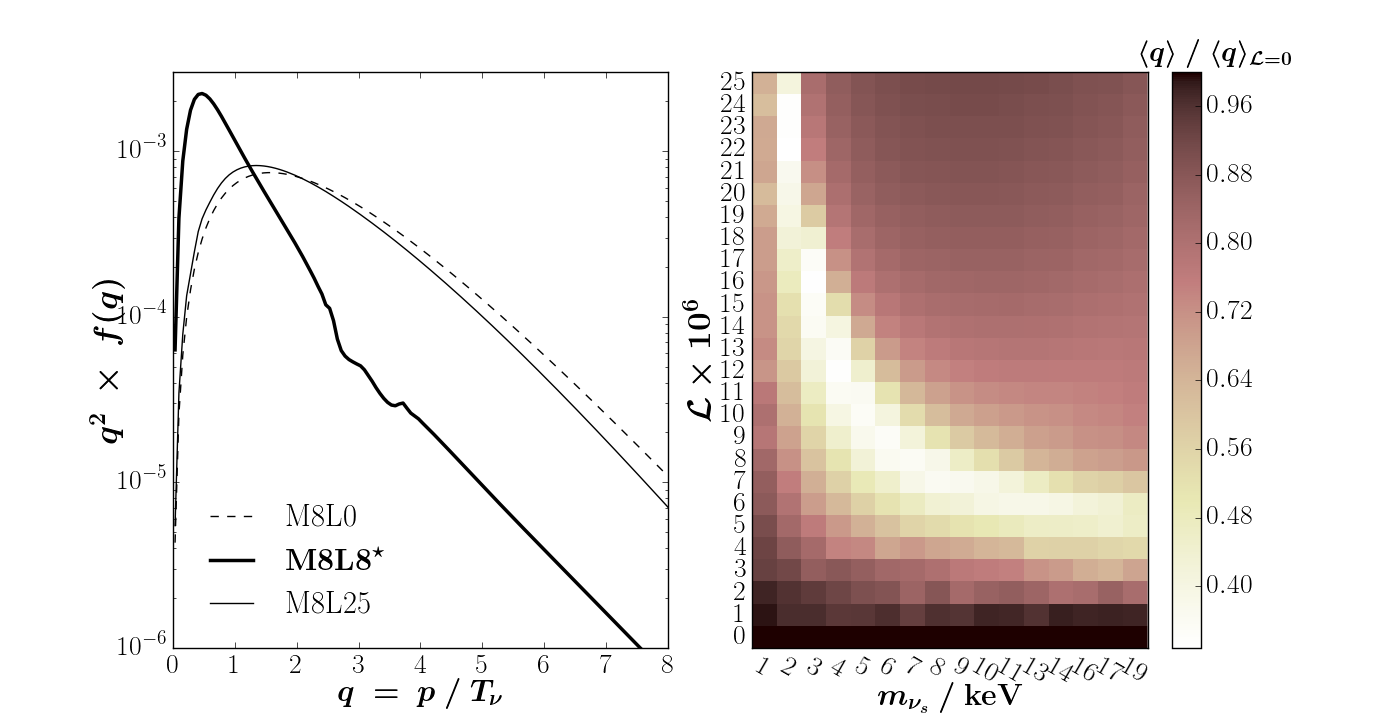}
\caption{ \textbf{Left:} Momentum distribution functions for 2 RPSNs with
  $m_{\nu_s}^{\mathrm{rp}} = 8~\rm{keV}$ in addition to the NRP
  $\mathcal{L}=0$ case shown as the grey dashed line. The coolest distribution
  function, occuring for an asymmetry of $\mathcal{L}^{\star} = 8 \times 10^{-6}$,
  is shown in thick solid black and labeled M8L8$^{\star}$. An asymmetry of
  $\mathcal{L} = 2.5 \times 10^{-5}$, labeled M8L25, results in higher momenta
  shown in thin solid black. \textbf{Right:} Average momentum
  $\langle q \rangle / m_{\nu_s}$ of the RPSN distribution functions computed
  in \cite{LaineMSM} normalised to the NRP $\mathcal{L}=0$ case for each mass
  (bottom-most row). Lepton asymmetries are in units of
  $\mathcal L = \left( n_{\nu_e} - n_{\bar{\nu}_{e}} \right) / s$. For each
  mass, the value of $\mathcal{L}^{\star}$ yielding the coolest distribution
  function is easily identifiable as the bright stripe.}
\label{fig:M8L8}
\end{center}
\end{figure*}

The right panel of Fig.~\ref{fig:M8L8} displays the value of $\langle q
\rangle / m_{\nu_s}$ for RPSN distribution functions computed in
\cite{LaineMSM,Ghiglieri:2015jua}. The coolest distribution functions occur for given values of $\mathcal{L}$ and $m_{\nu_s}^{\rm{rp}}$, which we denote $\mathcal{L}^{\star} (m_{\nu_s})$. For larger asymmetries than $\mathcal{L}^{\star}$ for a given mass, the resonantly boosted forward scattering occurs later than the QCD phase transition, which yield quasi-Fermi populated momenta states (with weaker mixing angles). The left panel of Fig.~\ref{fig:M8L8} features the distribution functions of several leptonic asymmetries for a $m_{\nu_s}^{\rm{rp}} = 8~\rm{keV}$ RPSN.

We run 8 hydrodynamics simulations for the following RPSN models: M3L16$^\star$, M4L12$^\star$, M6L6, M6L9$^\star$, M7L8$^\star$, M8L4,
M8L8$^\star$ and M13L6$^\star$, where M is the RPSN mass in keV and L the asymmetry parameter in units of $10^{-6}$. Their linear matter power spectra
at $z=30$, input to the non-linear code  \textsf{Gadget},  are computed from the distribution functions using the \textsf{CLASS} software.  


\section{Constraints on NCDM mass from Ly-$\alpha$ forests}
\label{sec:results}
\subsection{Mapping between C+WDM and RPSN as cool DM}
\label{subsec:mapping}

\begin{figure}
\begin{center}
\includegraphics[width=0.8\columnwidth]{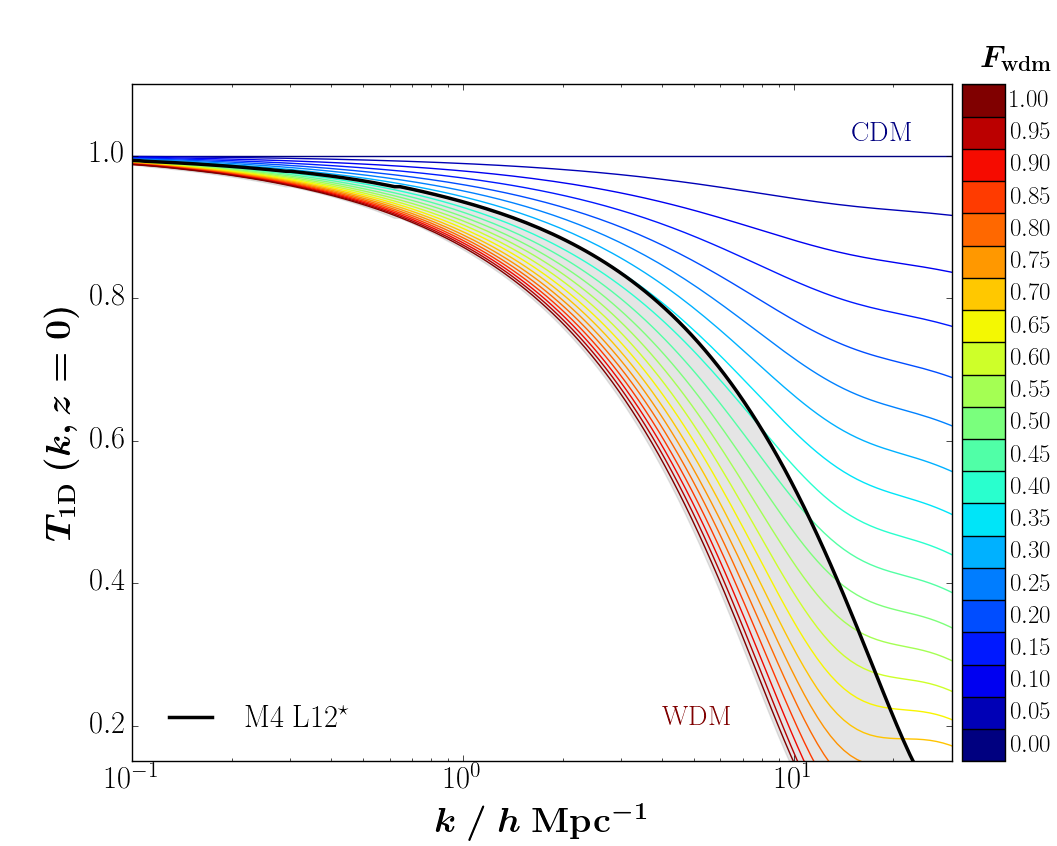}
\caption{ 1D linear transfer function for the total matter at $z=0$ obtained with the \textsf{CLASS} software for a DM made of a $4~\rm{keV}$ sterile neutrino. The color encodes the warm-to-total fraction $0 \leq F_{\rm{wdm}} \leq 1$, which ranges from the warmest (pure WDM) to the coolest (CDM) cases. Considering sterile neutrino pure cool dark matter, the transfer functions for the $0 \leq \mathcal{L} \leq \mathcal{L}^\star$ and $\mathcal{L}^\star \leq \mathcal{L} \leq \mathcal{L}^{\mathrm{max}}$ RPSN are all contained within the shaded grey region, bounded by the warmest and coolest models, respectively $\mathcal{L}=0$ (NRP in dark red) and $\mathcal{L}^\star = 1.2 \times 10^{-5}$ for $m_{\nu_s} = 4~\rm{keV}$ in thick black. }
\label{fig:M4t1d}
\end{center}
\end{figure}

As we  discussed in the previous section, there are two ways one can make the dark matter distribution cooler than the Dodelson-Widrow pure WDM case. When one considers a sterile neutrino of mass $m_{\nu_s}$, one can thus explore the following possibilities:
\begin{itemize}
\item[$\bullet$] the sterile neutrino constitutes the entirety of dark matter ($F_{\rm{wdm}} = 1$) and is produced in absence of a net leptonic asymmetry ($\mathcal{L} = 0$);
\item[$\bullet$] the sterile neutrino constitutes $0 \leq F_{\rm{wdm}} < 1$ of the total dark matter and is produced in absence of a net leptonic asymmetry ($\mathcal{L} = 0$);
\item[$\bullet$] the sterile neutrino constitutes the entirety of dark matter ($F_{\rm{wdm}} = 1$) and is produced in presence of a net leptonic asymmetry in the early Universe $\mathcal{L} > 0$;
\item[$\bullet$] the sterile neutrino constitutes $0 \leq F_{\rm{wdm}} < 1$  of the total dark matter and is produced in presence of a net leptonic asymmetry in the early Universe $\mathcal{L} > 0$.
\end{itemize}

In \cite{Baur16}, we  explored the first of these 4 listed scenarios and have concluded that to be consistent with Ly-$\alpha$ forest data from SDSS-III, the pure WDM sterile neutrino has to be more massive than $24.4~\rm{keV}$ with 95\% likelihood\footnote{the lower bound is relaxed to $16.0~\rm{keV}$ when adding CMB data}. This lower bound is at 12$\sigma$ tension with the upper bound issued by X-ray data, set at $4~\rm{keV}$. Scenario 1 has thus been disfavored. Search for keV sterile neutrino DM has shifted to scenarios 2 or 3, which we investigate in this paper. We have run 36 hydrodynamical simulations with a resolution of $2 \times 3072^3$ particles in a $(100~h^{-1}\rm{Mpc})^3$ co-volume, 28 of which explored C+WDM models ( Sec.~\ref{subsec:cwdm}) while the remaining 8 were dedicated to RPSN as pure cool dark matter (Sec.~\ref{subsec:rpsn}). To extend the sterile neutrino parameter space we sample, we use the 28 C+WDM configurations in addition to the 8 hydrodynamical simulations modelling RPSN. We take advantage of the fact that both yield similar transfer functions up to some $k$ scale to map out a correspondance between ($\mathcal{L}>0$, $F_{\rm{wdm}}=1$) and ($\mathcal{L}=0$, $0 \leq F_{\rm{wdm}} < 1$) assuming particles of the same mass $m_{\nu_s}$.
 
\begin{figure}
\begin{center}
\includegraphics[width=\columnwidth]{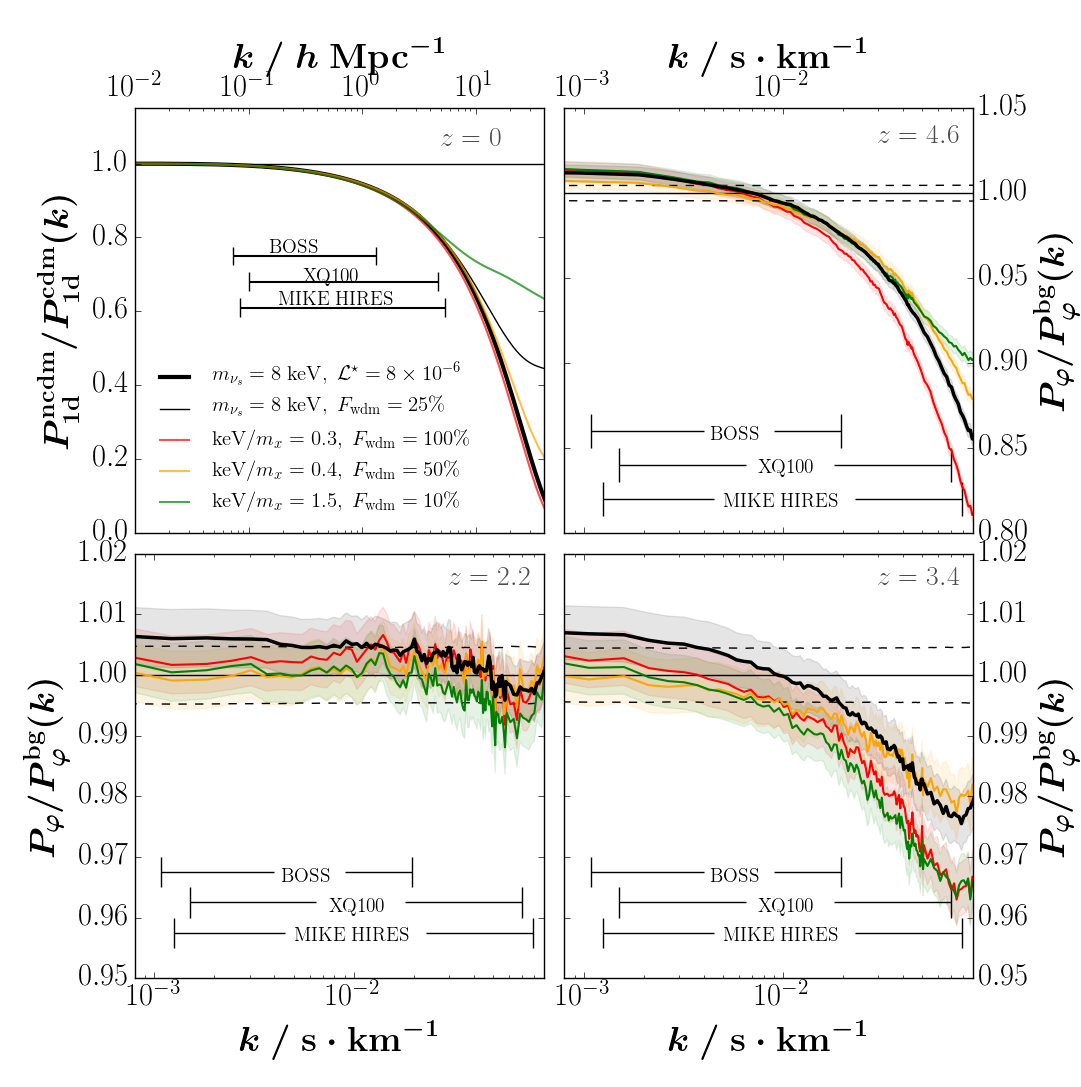}
\caption{Power spectra of the ($m_{\nu_s}/\mathrm{keV}=8$, $\mathcal{L}_6=8$) simulation normalized to the \textit{best guess} configuration, along with the ($\mathrm{keV}/m_x=0.3$, $F_{\rm{wdm}}=100\%$), ($\mathrm{keV}/m_x=0.4$, $F_{\rm{wdm}}=50\%$) and ($\mathrm{keV}/m_x=1.5$, $F_{\rm{wdm}}=10\%$) models. {Top Left:} 1D Linear matter power spectra ratio produced by \textsf{CLASS}. {Clockwise from Top Right:} Flux power spectra ratio produced by our hydrodynamical simulations at redshifts $z=4.6, 3.4$ and $2.2$. Shades encode simulation uncertainties (dotted lines for \textit{best guess}).}
\label{fig:flux}
\end{center}
\end{figure}

Fig.~\ref{fig:M4t1d} illustrates the correspondance between the coolest RPSN model of 4 keV ($\mathcal{L}^\star = 1.2 \times 10^{-5}$) and the 4 keV neutrino produced in absence of a lepton asymmetry that constitutes $\sim 35\%$ of the total dark matter. The correspondance is obtained with a least-square method out to $k_{\rm max} = 1.35\,h^{-1}{\rm Mpc}$ on the linear transfer function. Because Ly-$\alpha$ forests are a unidimensional probe for the matter distribution, we perform our mapping on the 1D transfer function, obtained using Eq.~\ref{eq:1D}. This $\mathcal{L}-F_{\mathrm{wdm}}$ mapping enables us to convert our bounds on ($m_{\nu_s}$, $F_{\rm{wdm}}$) obtained with our 28 C+WDM simulations to bounds on ($m_{\nu_s}$, $\mathcal{L}$). 

For illustration purposes, the top left panel in Fig.~\ref{fig:flux} shows the linear power spectra of the M8L8$^\star$ (in thick black) along with the closest matching linear matter $T_{\mathrm{1d}}(k)$ of a C+WDM model assuming $m_{\nu_s} = 8~\rm{keV}$, which occurs for a warm-to-total DM fraction of $F_{\rm{wdm}} = 25\%$ (in thin black). As illustrated in Fig.~\ref{fig:M4t1d}, this linear transfer function correspondance is adequate up to some $k$ scale, beyond which the corresponding C+WDM model $T_{\mathrm{1d}}(k)$ breaks away from its comparative RPSN model to an asymptotical plateau ($T_{\mathrm{1d}}(k \rightarrow \infty) \propto (1-F_{\mathrm{wdm}}) \geq 0$). For most values of $m_{\nu_s}$ explored in this work, this breakaway $k$ is beyond the scales probed by our Ly-$\alpha$ forest data set, which we've materialized on Fig.~\ref{fig:flux}. We illustrate the negligible impact of differences in the linear 1D transfer function beyond the breakaway scale by overplotting three C+WDM models that exhibit similar $T_{\mathrm{1d}}$ on large scales.  The differences in the non-linear regime measured by the flux power spectra are within the statistical uncertainties of the simulations, almost an order of magnitude smaller than data uncertainties on similar scales.\\

While we derive constraints in the sterile neutrino parameter space through the mapping from our C+WDM grid described above, we also use the eight exact RPSN hydrodynamical simulations to determine the accuracy of our mapping procedure. To this end, we compare the $\chi^2$ obtained 
using the exact (non-linear) RPSN simulation with that of its corresponding C+WDM model obtained with our mapping procedure in the linear regime (see Fig.~\ref{fig:flux} for illustration with M8L8$^{\star}$).
This cross-check is done for both sets of data, SDSS/BOSS alone ('SDSS') and combined with VLT/XShooter, Keck/HIRES and LCO/MIKE ('SDSS+XQ+HR'). We observe a systematic shift at the level of $0.2\,\sigma$ on average in the first case, and of $0.5\,\sigma$ in the second, with the RPSN simulation showing a smaller $\chi^2$ (better agreement with the data) than its C+WDM match. The better agreement for SDSS only is consistent with a better match of the transfer function on large scales. The results we present hereafter for the RPSN models are corrected for this systematic shift. 

\subsection{Constraints on C+WDM mass and fraction}
\label{subsec:res_cwdm}

The probability distribution in the $(F_{\rm{wdm}},m_x)$ plane  being strongly non-Gaussian, a Taylor expansion in either of these two parameters would not provide  accurate results. 
We therefore extend the method described in~\cite{Borde2014, Palanque2015a, Palanque2015b, Baur16} in the following way. We use the likelihood of previous work, based on a second-order Taylor expansion, to capture the dependence  of the Ly-$\alpha$ flux power spectrum with the cosmological and astrophysical variables of Table~\ref{tab:params}, and to model identified nuisance parameters accounting for IGM thermal state modeling, re-ionization redshift, spectrometer noise, and simulation uncertainties. 
To capture the dependence with $F_{\rm{wdm}}$ and $m_x$, we produced a grid  of 28 C+WDM simulations with non-zero $(F_{\rm{wdm}},m_x)$ while all other parameters are set to their \textit{best-guess} value in Table~\ref{tab:params}. 
For each of the C+WDM simulations, we  compute a $\chi^2$  with respect to our $35 \times 12$ $P_\varphi(k, z)$ data points from the BOSS DR9 described in Sec.~\ref{subsec:data}, assuming a Gaussian distribution for the Hubble parameter, $h=0.673 \pm 0.010$ (Planck 2015), and minimizing over all the parameters of the likelihood described above. We interpolate within this grid to predict the probability distribution at any point in the $(F_{\rm{wdm}},m_x)$ plane.

\begin{figure}
\begin{center}
  \includegraphics[width = 0.55\textwidth]{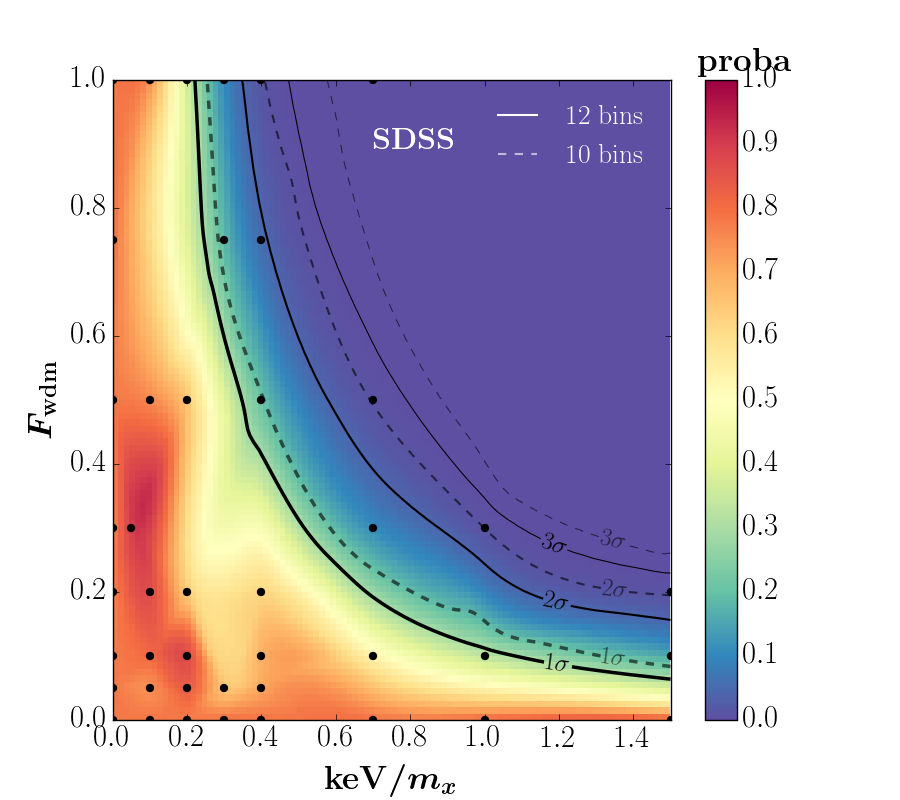}~%
  \includegraphics[width = 0.55\textwidth]{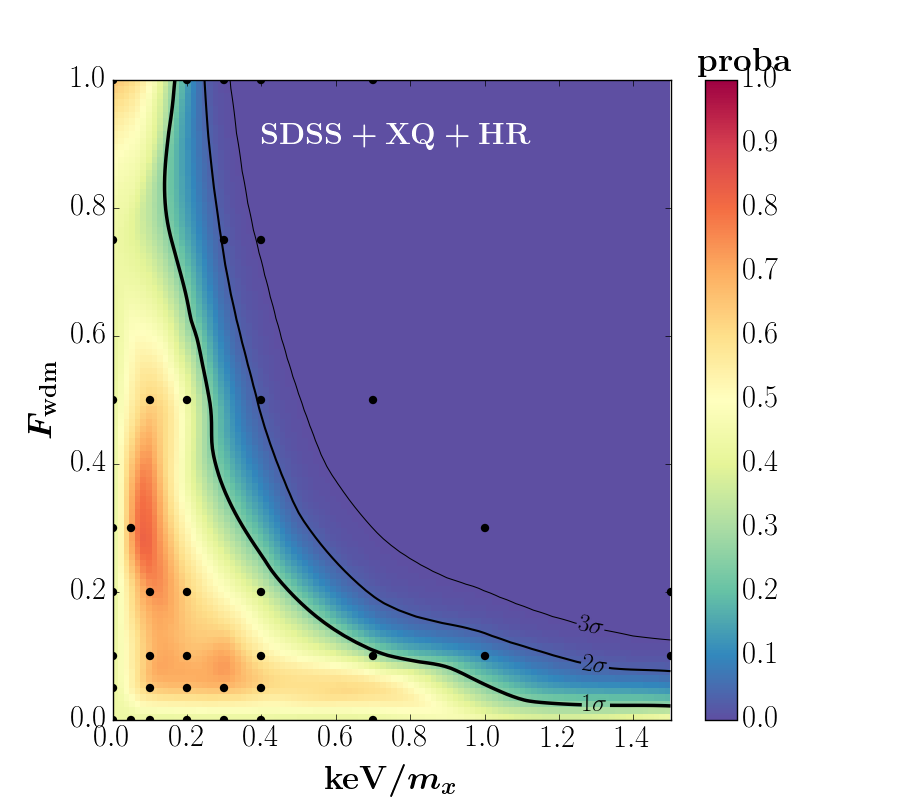}
  \caption{Our set of C+WDM hydrodynamical simulations mapped as black dots on
    our grid of ($\mathrm{keV}/m_x$, $F_{\rm{wdm}}$). Note the points
    corresponding to either $F_{\rm{wdm}}=0$ or $\mathrm{keV}/m_x=0$ all
    correspond to the \textit{best guess} model that assumes a CDM
    cosmology. The color scheme reflects the probability function defined in
    Eq.~\ref{eq:proba}. {Left:} bounds based on the SDSS/BOSS data
    only. Solid curves are 1, 2 and 3$\sigma$ CL contours using all 12
    redshift bins, while the dashed curves materialize the contours when
    excluding the 2 highest redshift bins (z=4.2 and 4.4) in the likelihood. {Right:}
    combined SDSS (all 12 redshift bins) + XQ + HR data set. Non-monotonic IGM thermal histories suggested
    in~\protect\cite{warmIGM} are not included in the marginalization.}
\label{fig:Chi2_cwdm}
\end{center}
\end{figure}

We identify the minimal value of $\chi^2$ using the \textsf{MINUIT} package~\cite{Minuit}, letting all parameters free. We set a confidence level (CL) on any parameter $\theta_i$ (out of $n$) by minimizing the $\chi^2$ function on all remaining $n-1$ parameters for each scanned value of $\theta_i$. To set confidence levels on a hypersurface of 2 parameters ($\theta_i$, $\theta_j$), the $\chi^2$ minimization is performed on the $n-2$ remaining parameters. Assuming all experimental errors are normally distributed, 
\begin{equation}
\mathrm{CL}(\theta_i, \theta_j, ..., \theta_n) = 1 - \int_{\Delta \chi^2 (\theta_i, \theta_j, ..., \theta_n)}^{\infty} \mathrm{d} x ~ f_{N_{\mathrm{dof}}}(x)
\label{eq:CL}
\end{equation}

\begin{equation}
f_{N_{\mathrm{dof}}}(x) ~ = ~ \frac{e^{-x/2} ~ x^{\frac{N_{\mathrm{dof}}}{2}-1}}{\sqrt{2^{N_{\mathrm{dof}}}} ~ \Gamma (N_{\mathrm{dof}}/2)}
\label{eq:proba}
\end{equation} where $\Gamma (z) = \displaystyle \int_{0}^{\infty} \mathrm{d}x~ x^{z-1} e^{-x}$ is the gamma function. $1\sigma$, $2\sigma$ and $3\sigma$ confidence levels we quote on parameter $\theta_i$ or ($\theta_i$, $\theta_j$) correspond to a $\chi^2$ value with respect to the minimum value of $\Delta \chi^2 (\theta_i) = 1, 4$ and 9 and $\Delta \chi^2 (\theta_i, \theta_j) = 2.30, 6.18$ and 11.83 respectively. 

Fig.~\ref{fig:Chi2_cwdm} displays a cubic-interpolated map of the $\Delta \chi^2$ probability in the ($\mathrm{1\,keV}/m_x$, $F_{\rm{wdm}}$) plane using the 28 C+WDM flux power spectra computed using our hydrodynamical simulations. The $\chi^2$ values we obtain along the $F=1$ axis are in excellent agreement with the ones we reported in \cite{Baur16} and \cite{Yeche17} for 95\% confidence level limits in the pure WDM case, which led to $m_x \geq 4.09~\rm{keV}$ and $m_{\nu_s}^{\rm{nrp}} \geq 24.4~\rm{keV}$ using our SDSS Ly-$\alpha$ sample only and to $m_x \geq 4.65~\rm{keV}$ and $m_{\nu_s}^{\rm{nrp}} \geq 28.8~\rm{keV}$ using in addition the XQ-100, HIRES and MIKE data. The bounds shown in Fig.~\ref{fig:Chi2_cwdm} at  $2\sigma$ are of course weaker than the ones reported previously since they are here computed for a full two-dimensional analysis in the ($\mathrm{1\,keV}/m_x$, $F_{\rm{wdm}}$) plane,  thus for $\Delta\chi^2 = 6.18$ instead of 4. 

Independently of the redshift range considered for  SDSS   and of the use or not of the higher-resolution data, the   contours in the ($\mathrm{1\,keV}/m_x$, $F_{\rm{wdm}}$) plane
can  be approximated  by $F_{\rm{wdm}} = \alpha(1\,{\rm keV}/m_x)^\beta$, with  $\beta = -1.37$ in all cases. For the 95\% C.L. limits, the normalization is $\alpha=0.24$ for the full (12-bin) SDSS data,  and $\alpha=0.14$ for SDSS+XQ+HR. For the $3~\sigma$ limits,  the normalization is $\alpha=0.35$ for  SDSS  and $\alpha=0.20$ for SDSS+XQ+HR.
Therefore, thermally decoupled relics as light as $m_x \geq 0.7~\rm{keV}$ are consistent with Ly-$\alpha$+$H_0$ data (95\% CL) if they constitute 15\% of the total dark matter or less. The contribution of  warm to total dark matter is reduced to $\sim 10\%$ when including  higher-resolution data.  
 The bounds are derived under the assumption that the IGM thermal history can be modeled with 5 parameters: a broken power law for $T_0$ and a simple power law for  $\gamma$. Non-monotonic thermal histories as discussed in~\cite{warmIGM}, for instance, are not included in the marginalization. Although the impact of such a hypothesis is small for SDSS-only bounds, which can therefore be considered as conservative lower bounds, it can lead to looser limits in the case of SDSS+XQ+HR.  

In the pure WDM case, we showed that the highest two redshift bins of the SDSS data significantly tightened the 95\% CL limit despite their low statistical significance~\cite{Baur16}. The same is true here for the study in the full ($\mathrm{1\,keV}/m_x$, $F_{\rm{wdm}}$) plane. Considering  the lowest ten redshift bins only (i.e., redshifts in $2.1<z<4.1$) loosens the  bound by about 25\% on $\mathrm{1\,keV}/m_x$ for a given $F_{\rm{wdm}}$.  The $\chi^2$ of the best fit increases by 71.7 when including the 70 SDSS data points at $z>4.1$, indicating that the highest two redshift bins are consistent with the bins at lower redshift. 

\begin{table}[h]
	\begin{center}
		\begin{tabular}{ccc}
			\textbf{Parameter} & \textbf{SDSS} & \textbf{SDSS+XQ+HR}\\[2pt]
			\hline \\[-10pt]
			$\sigma_8$ & $0.855\pm 0.020$ & $0.815\pm 0.020$\\[2pt]
			$n_s$& $0.935\pm0.010$ & $0.950 \pm 0.010$\\[2pt]
			$T_0\;(z=3)$  \scriptsize{(K)} & $9600\pm4000$ & $14500 \pm 3000$  \\[2pt]
			$\gamma$ & $0.9\pm 0.2$ & $0.9 \pm 0.2$\\[2pt]
			$\eta^{T_0}\;(z<3)$ & $-2.7\pm 0.7$ & $-1.9 \pm 0.4$\\[2pt]
			$\eta^{T_0}\;(z>3)$ & $-4.1\pm 1.4$ & $-2.1\pm 0.6$\\[2pt]
			$\eta^{\gamma}$ & $0.7\pm 0.5$ & $-0.3\pm 0.4$ \\[2pt]
			\hline \\[-10pt]
		\end{tabular}
	\end{center}
	\caption{Best-fit values and 68\% C.L. of the most relevant parameters. We take $\mathrm{1\,keV}/m_x=0.05$, $F_{\rm{wdm}}=0.30$ as the best-fit C+WDM model, although values for the  parameters of the table change by $<1\;\sigma$ along the degeneracy $F_{\rm{wdm}} = \alpha(1\,{\rm keV}/m_x)^\beta$  curve as well as from a pure WDM fit. }
	\label{tab:bestfit_values}
\end{table}
The best-fit parameters for either a pure WDM or a C+WDM scenario, shown in Table~\ref{tab:bestfit_values}, are all in excellent agreement with our previous WDM analyses \cite{Baur16,Yeche17}.  The IGM temperature history is compatible with other recent estimates~\cite{IrsicWDM} and the cosmological parameters are compatible with 
 the latest Planck results, except for a $\sim 2\sigma$ tension on $n_s$ when fitting BOSS data alone.  As was shown in~\cite{Baur16}, the lower preferred value of $n_s$ in BOSS Ly$\alpha$ data compared to CMB has an impact on the constraint one can set on the mass of a pure WDM particles. The use of  BOSS Ly$\alpha$ data alone, or, equivalently, allowing for a running of $n_s$ that accommodates for the different values of $n_s$ on large (CMB regime) and small (Ly$\alpha$ regime) scales, leads to tighter constrains than when fitting  BOSS and Planck data together in the absence of running. A similar effect is true here. Approximating the 95\% CL contour by  $F = \alpha(1\,{\rm keV}/m_x)^\beta$ as we did above, we obtain a constraint on  $m_x$ that is looser by about 35\% for fixed $F_{\rm{wdm}}$ when adding  Planck to BOSS data. The situation is different with the extended SDSS+XQ+HR data, for which it was shown in~\cite{Yeche17} that the tension on $n_s$ was mostly resolved. We thus expect similar constraints on $m_x$ whether or not we include  CMB data in addition to this extended set.   For fixed $F_{\rm{wdm}}$, the 95\% CL constraint on $m_x$ indeed shifts by less than 12\% between the two configurations (extended Ly$\alpha$ alone or with the addition of Planck).

\subsection{Constraints on RPSN mass}
\label{subsec:res_rpsn}

\begin{figure}
\begin{center}
  \includegraphics[width = 0.55\textwidth]{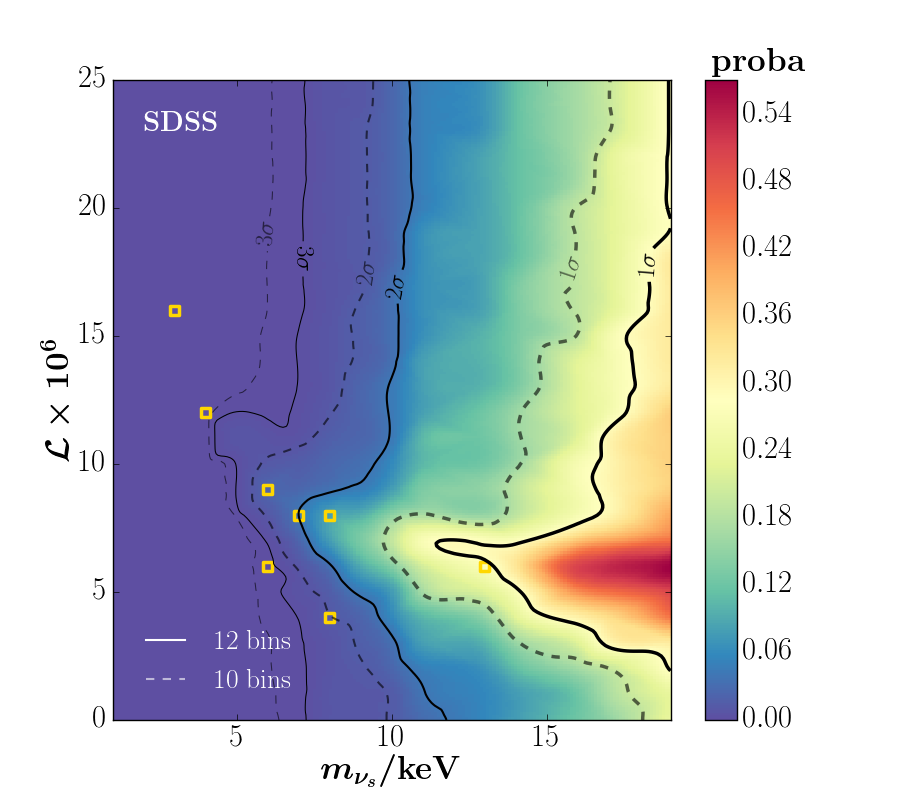}~%
  \includegraphics[width = 0.55\textwidth]{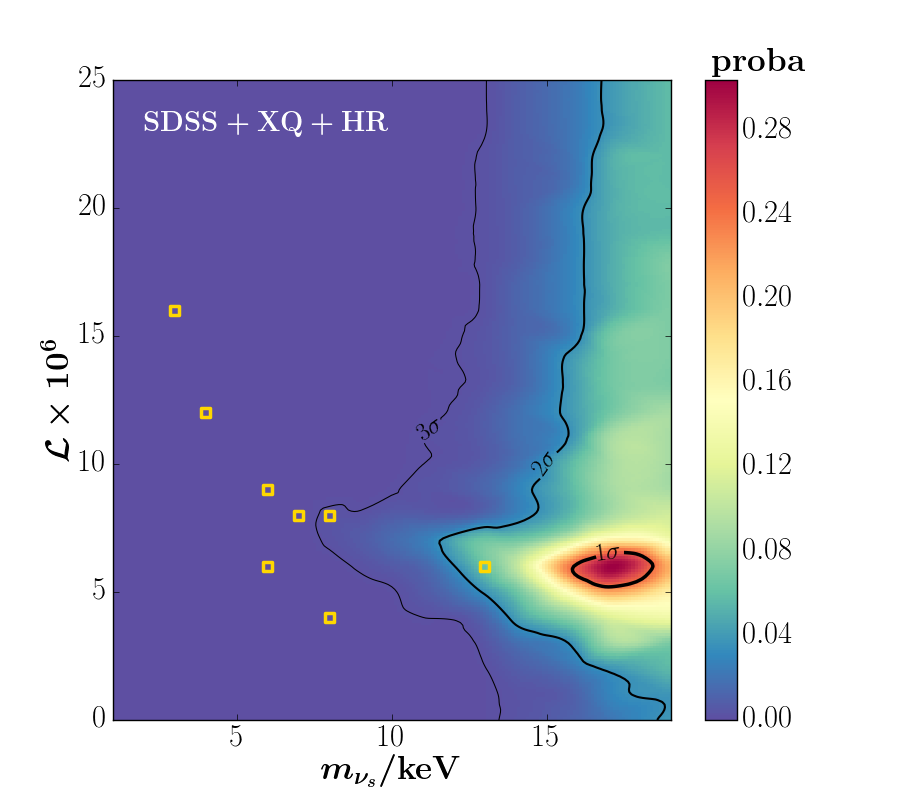}
  \caption{Constraints on ($m_{\nu_s}$, $\mathcal{L}$) obtained by the mapping
    described in Sec.~\ref{subsec:mapping}. Similar color scheme as in
    Fig.~\ref{fig:Chi2_cwdm}. Gold squares map the 8 RPSN models we ran with our hydrodynamics simulations. \textbf{Left:} SDSS/BOSS data only. Solid curves are
    1, 2 and $3\sigma$ CL using all 12 redshift bins, while the dashed curves
    materialize the contours when excluding the 2 highest redshift bins (z=4.2
    and 4.4) in the likelihood. \textbf{Right:} Combined SDSS (all 12 redshift bins) + XQ +
    HR data sets.  Non-monotonic IGM thermal histories suggested
    in~\protect\cite{warmIGM} are not included in the marginalization.
    }
\label{fig:RPSN_ML}
\end{center}
\end{figure}

\begin{figure}
\begin{center}
\includegraphics[width=\columnwidth]{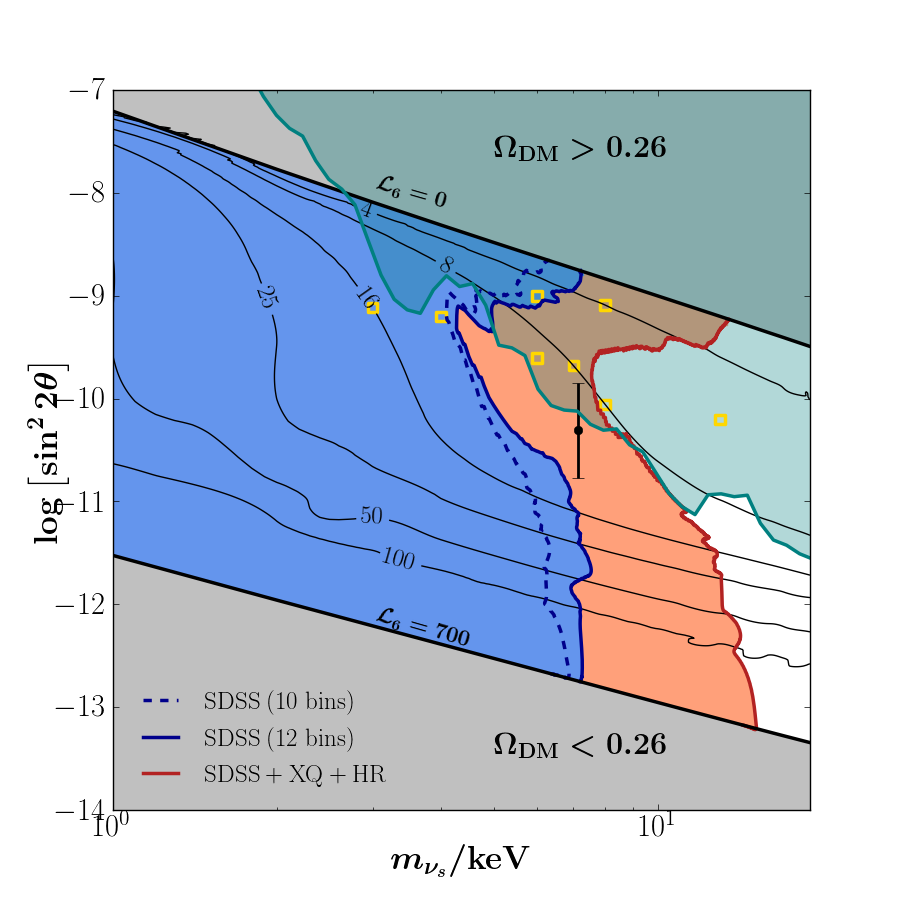}
\caption{Constraints from Ly-$\alpha$ forest in the RPSN ($m_{\nu_s}$,
  $\sin^2 2 \theta$) parameter space. The iso-$\mathcal{L}$ contours are displayed in black along with the corresponding value of $\mathcal{L}_6$. Gold squares indicate the set of parameters for which we computed the $P_\varphi (k)$ by solving the non-linear hydrodynamics. The black dot with error bar denotes the right-handed interpretation of the $3.55 ~\mathrm{keV}$ X-ray line in the stacked spectra of galaxy clusters, for which we used  $m_{\nu_s} = 7.14 \pm 0.07 ~\mathrm{keV}$ and $\sin^2 2 \theta = 4.9^{+ 1.3}_{-1.6} \times 10^{-11}$ as reported in \cite{Boyarsky14}. The blue  (\textit{resp.} red) shade encompasses  models excluded by over $3\sigma$ by the SDSS-only  (\textit{resp.} SDSS + XQ + HR) Ly-$\alpha$ forest  power spectrum. The absence of monochromatic X-ray lines (apart from the  $3.55~\mathrm{keV}$ signal) translate into upper bounds in $\sin^2 2 \theta (m_{\nu_s})$: the green shade are models inconsistent beyond $3\sigma$ with a compilation of X-ray data from the Milky Way, Andromeda and other galaxies.}
\label{fig:RPSN_MT}
\end{center}
\end{figure}

We convert the bounds on ($m_x$, $F_{\mathrm{wdm}}$) established in the previous
subsection  into bounds on ($m_{\nu_s}$, $\mathcal{L}$) using the mapping
procedure described in Sec.~\ref{subsec:mapping}, corrected for systematic shifts using the
$P_{\varphi}$ from our 8 RPSN hydrodynamical simulations. Fig.~\ref{fig:RPSN_ML}
displays the 1$\sigma$, 2$\sigma$ and 3$\sigma$ confidence level contours in the
($m_{\nu_s}$, $\mathcal{L}$) plane using our SDSS/BOSS Ly-$\alpha$
data  (left panel) or the full SDSS+XQ+HR  data (right
panel). The tension in standard deviations with these two data sets is reported in Table~\ref{tab:RPSNsigma} for the  relevant RPSN models for which we ran hydrodynamical simulations.
A lepton asymmetry during the era of RPSN production in the early Universe
boosts the oscillation frequency from active to sterile neutrinos, thus
enabling ample production of dark matter sterile neutrinos with weaker mixing
angles $\theta$. We run a Boltzmann code that computes the mixing angle
for a given dark matter density, mass and lepton asymmetry
parameter. Fig.~\ref{fig:RPSN_MT} displays the quantity $\sin^2 2 \theta$ as a
function of $m_{\nu_s}$ assuming $\Omega_{\mathrm{dm}} h^2 = 0.26142 \times
0.675^2\simeq 0.119$ and values of the lepton asymmetry parameter in the range $0 \leq \mathcal{L} \leq 7 \times 10^{-4}$. The black lines in Fig.~\ref{fig:RPSN_MT} display the relation between  mass and mixing angle  for eight values of the primordial lepton asymmetry shown along each black curve (in  units of $10^{-6}$). The $\mathcal{L}_6 = 0$ thick line corresponds to  non-resonant production. Values above $\mathcal{L}\gtrsim 10^{-3}$ are inconsistent with Big Bang nucleosynthesis (BBN). For $\sin^2 2 \theta \gtrsim 10^{-7}$, dark matter is overproduced. The region excluded at the  $3\sigma$ CL by  SDSS/BOSS (\textit{resp.} by SDSS + XQ + HR) data  is shaded in blue  (\textit{resp.} in red).  The dashed blue line indicates the limit using the 10 lowest redshift bins only, i.e., excluding $z=4.2$ and 4.4. 

\begin{table}[h]
	\begin{center}
		\begin{tabular}{lcc}
			\multicolumn{1}{c}{\textbf{RPSN model}} & \textbf{SDSS} & \textbf{SDSS+XQ+HR}\\[2pt]
			\hline \\[-10pt]
			M4L12$^\star$ & $3.06 ~\sigma$ & $ >4\sigma$\\[2pt]
			M6L6 & $3.13 ~\sigma$ & $ >4\sigma$\\[2pt]
			M6L9$^\star$ & $2.0 ~\sigma$ & $3.3 ~\sigma$\\[2pt]
			M7L8$^\star$ & $1.9 ~\sigma$ & $3.1 ~\sigma$\\[2pt]
			M8L4 & $2.7 ~\sigma$ & $> 4 ~\sigma$\\[2pt]
			M8L8$^\star$ & $1.5 ~\sigma$ & $2.5 ~\sigma$\\[2pt]
			\hline \\[-10pt]
		\end{tabular}
	\end{center}
	\caption{Tension with respect to measured  $P_{\varphi}(k)$ for  the hydrodynamical simulations that fall in the range between 1 and  4~$\sigma$. RPSN configuration is given in column 1,    comparison to `SDSS only' in column 2  and to `SDSS+XQ+HR' in column 3.}
	\label{tab:RPSNsigma}
\end{table}

As expected from the white stripe visible on the right panel of Fig.~\ref{fig:M8L8}, the coolest RPSN models, which occur for $\mathcal{L} = \mathcal{L}^{\star}(m_{\nu_s})$, feature the longest free-streaming length and are more consistent with Ly-$\alpha$ forest data than other values of $\mathcal{L}$. This is visible on Fig.~\ref{fig:RPSN_ML} as a horn-like valley in the $\chi^2$ maps, 
 which extends to sterile neutrino masses around $\sim 7~\mathrm{keV}$ in the right-hand panel. 
This area  is of particular interest since it matches the range of masses and mixing angles for which the $3.55~\mathrm{keV}$ X-ray  signal reported in \cite{Bulbul14,Boyarsky14,Boyarsky2015}  can be interpreted as photons emitted by the decay of a $7.1~\mathrm{keV}$ right-handed neutrino. Although this region exhibits a $\sim3\sigma$ tension with the SDSS+XQ+HR Ly-$\alpha$ data, two  caveats should be considered. First,  IGM thermal histories impact the small scales ($0.02 ~\leq~ k / s~\mathrm{km}^{-1}~ \leq~ 0.07$) probed by these high-resolution data. Although we marginalize over 5 parameters to describe the thermal history  (as explained above), more general models (non-monotonic temperature evolution for instance) could loosen our constraint. Second, the flux power spectrum exhibits large gradients with respect to the RPSN parameters around the ``horn" region,  where the interpolation procedure is thus more delicate.
Therefore, because of the interest of this region for RPSN constraints, we located our eight RPSN simulations in that area: six correspond to the coolest models for their mass  (M3L16$^\star$, M4L12$^\star$, M6L9$^\star$, M7L8$^\star$, M8L8$^\star$ and M13L6$^\star$), the remaining two (M6L6 and M8L4) being slightly warmer than their corresponding counterparts at the same mass (M6L9$^\star$ and M8L8$^\star$). 
The results presented in Table~\ref{tab:RPSNsigma} show that the horn is a real feature,  although the exact location of its boundaries might require additional hydrodynamical simulations to assess. 
Hence the shape of the blue and red contours on Fig.~\ref{fig:RPSN_MT} may be less accurate  in the regions around the 6 bottom-most gold squares that correspond to our coolest RPSN models. 
The neutrino decay origin of the $3.55~\mathrm{keV}$ X-ray line, shown as the black dot with error bars, is located in this  region. 

For the reasons just stated, we suggest scanning the  area  around  $m_{\nu_s} = 7.1  ~\mathrm{keV}$ and $\sin^2 2 \theta = 4.9 \times 10^{-11}$ with a set of dedicated hydrodynamical simulations in order to properly account for the strong dependence of the power spectrum on model parameters in that region. These simulations should also implement the different IGM thermal histories prognosticated in \cite{warmIGM}. We leave this for future work.


\section{Conclusion}
\label{sec:conclusion}
We used the SDSS-III/BOSS DR9 Ly-$\alpha$ forest data to constrain 
warm dark matter models. In the 
previous study~\citep{Baur16}, these data were 
used  to put bounds on the mass of warm dark matter in the form of
\emph{thermal relics}.
In this paper,  we extend the previous results to two important 
classes of non-thermal WDM models: cold-plus-warm dark matter and
sterile neutrinos produced via mixing with active neutrinos in the
presence of a net lepton asymmetry (known as \emph{resonantly produced sterile
neutrinos}). While many works have used 
Ly-$\alpha$ forest  to constrain thermal relic WDM (and non-resonantly produced sterile 
neutrinos)~\cite{Hansen:2001zv,VLH08a,SMT08,VLH08b,VBH13,Yeche17,IrsicWDM,
Baur16}, in the current study, for the first time,  RPSN are constrained  by
running a set of hydrodynamical simulations with realistic initial power
spectra predicted by this scenario. 

The size of the SDSS/BOSS DR9 quasar dataset significantly reduces statistical uncertainties compared to previous SDSS-II data. The systematic uncertainties associated with feedback and IGM modeling, instrument noise \textit{etc}, now become comparable to the statistical ones. Uncertainties in the IGM thermal history, in particular, mostly affect the highest redshift bins. To provide both conservative (although weaker) and more ambitious (although more prone to systematics) bounds, we computed all limits with and without the two highest z = 4.2, 4.4 redshift bins.
These bounds can be additionally strengthened by 
including the higher-resolution XQ-100, MIKE and HIRES (XQ + HR) Lyman-$\alpha$ datasets that probe the power spectrum at
smaller scales and higher redshifts (see Fig.~\ref{fig:DR9fPS}). 
 In this regime, however, 
the flux power spectrum
exhibits a suppression on the smallest scales  that  a number of
astrophysical effects can lead to: Doppler broadening,
pressure smoothing, expansion of the filaments along the line of sight \textit{etc.}, \cite{VBH13,warmIGM}.  
Dark matter free-streaming  could also contribute to this suppression.\footnote{A possible way to identify the mechanism responsible for the small-scale suppression of the flux power spectrum is to measure the thermal history of the IGM 
  independently of  Ly-$\alpha$ forest data. To this end, a novel method ---
  \emph{Gaussian optical depth decomposition} --- was proposed in
\cite{Garzilli:2015bha}. It is based on the idea that for high-resolution
  spectra it is possible to identify individual absorption lines and to infer
  gas temperature directly via measurement of their \emph{broadening}. This
  project is currently underway. }
The detailed analysis of the influence of the thermal history on the flux
power spectrum at small scales is beyond the scope of this study. 
In this work, we  focus on thermal histories that can be modeled with 5 parameters: a broken power law for $T_0$ and a simple power law for $\gamma$. 

RPSNs lighter than $m_{\nu_s} \sim 3.5$~keV produce a Ly-$\alpha$ flux power spectrum inconsistent with the one measured from SDSS/BOSS at  more than $3\sigma$.   This bound tightens to $m_{\nu_s} \sim 7$~keV when 
including  the higher resolution data.
Heavier sterile neutrinos can be consistent with
 data if one assumes lepton asymmetries of about
$\mathcal L \sim 10^{-6} - 10^{-5}$ at the time of RPSN production.  Outside this range of lepton asymmetries, 
 both  limits tighten by about a factor~2. Models that would potentially explain the observed  $3.5~\mathrm{keV}$ line are compatible with  SDSS/BOSS 10-bin and 12-bin data at a level between  $2\sigma$ and $3\sigma$,  and with the extended SDSS+XQ+HR data at slightly above $\sim 3\sigma$.\footnote{Curiously, the 7~keV RPSN with the mixing
  angle within the range reported in~\cite{Boyarsky14,Bulbul14} produces the
  suppression of the flux power spectrum at small scales that would fit the
  XQ + HR data if the intergalactic medium is cold at redshifts $z \ge 5$~\cite{warmIGM}. }
We
notice that the distribution function of RPSN  for the coolest models (those in the  region of the white stripe of 
Fig.~\ref{fig:M8L8}) varies drastically with $m_{\nu_s}$ and $\mathcal L$. It is
therefore challenging to obtain precise constraints in this region of the RPSN
parameter space.   We therefore also run 8 hydrodynamical simulations with exact input for RPSN models, among which one corresponding to a 7 keV neutrino produced in a lepton asymmetry  $\mathcal{L} = 8 \times 10^{-6}$. This corresponds to the coolest, and thus most conservative, model for $m_{\nu_s}=7~\mathrm{keV}$. Its power spectrum   is consistent at $1.9~\sigma$ with BOSS data, and at $3.1~\sigma$ with BOSS + XQ + HR. 
We leave further  dedicated investigation of the region of interest
for future work.

The combined BOSS and XQ + HR Ly-$\alpha$ data sets are only compatible with a very restricted region in parameter space of RPSNs. This result is consistent with the conclusion of~\cite{CherryHoriuchi17} based on the clustering of SDSS galaxies.

Finally, in the case of C+WDM, we find that thermally decoupled relics as light as
$m_x \ge 0.7$~keV are consistent  at 95\% C.L. with SDSS Ly-$\alpha$ data if they
constitute less than 15\% of the total dark matter.  The bounds on $F_{\rm{wdm}}$ tighten to 10\% when  derived from the SDSS + XQ + HR data set. More generally, we find that the limits can be well approximated by  $F_{\rm{wdm}} = \alpha(1\,{\rm keV}/m_x)^\beta$, with $\beta\sim -1.37$.  At 95\% C.L., $\alpha = 0.24$ when using SDSS only and $\alpha=0.14$ with SDSS+XQ+HR. At $3~ \sigma$, $\alpha = 0.35$ for SDSS only and $0.20$ for SDSS+XQ+HR.

RPSNs require significant lepton asymmetry to be present in the primordial
plasma during the epoch of  dark matter
production~\cite{ShiFuller99,LaineMSM}. Such a lepton asymmetry can be
generated \textit{e.g.} in the $\nu$MSM~\cite{Shaposhnikov:2008pf}, minimal extension of the Standard Model
with 3 right-handed neutrinos that would explain both  the dark matter but also the 
baryon asymmetry of the Universe as well as neutrino
masses and oscillations~\cite{nuMSM_1,nuMSM_2,ARNPS_nus}. 

Sterile neutrino dark matter remains to this day an active area of
research. We focused on sterile neutrinos produced by the mixing with active
neutrinos, yet other production mechanisms can yield cool transfer functions
as well (see~\cite{Adhikari:2016bei} for a review). Our bounds do not apply to
these alternative production scenarios. Our work shows that Ly-$\alpha$ forest
constraints are consistent with those of other studies. Considering the
possibility that sterile neutrinos are a fraction of dark matter (just like
left-handed neutrinos are), any sterile neutrino mass is still allowed, given
that it constitutes at most $\sim 10\%$ of the total dark matter density and is
consistent with upper bounds from X-ray data.



\acknowledgments

We thank Matteo Viel for providing us high resolution quasar spectra from the MIKE and HIRES, as well as Volker Springel for making \texttt{GADGET-3} available to our team. JB thanks Alexander Merle and Kevork Abazajian for their helpful contributions and discussions on the resonant and non-resonant production mechanisms. This project  received funding from the European Research Council 
(ERC) under the European Union's Horizon 2020 research and innovation 
programme (GA 694896).
\\
We acknowledge PRACE (Partnership for Advanced Computing in Europe) for access to thin and xlarge nodes on the \textsf{Curie} cluster based in France at the TGCC (Tr\`es Grand Centre de Calcul) under allocation numbers 2010PA2777, 2014102371 and 2012071264. We also acknowledge the French national access to high-performance computing GENCI (Grand \'Equipement National de Calcul Intensif) for access to the \textsf{Curie} cluster under allocation t2016047706.
\\


\bibliographystyle{JHEP}
\bibliography{biblio}

\end{document}